\documentclass[journal]{IEEEtran}
\ifCLASSINFOpdf
  \usepackage[pdftex]{graphicx}
  \graphicspath{{./}}
  \DeclareGraphicsExtensions{.pdf,.jpeg,.jpg,.png}
\else
\fi
\usepackage{amsmath}
\usepackage{amsfonts}
\usepackage{amssymb}  
\usepackage[]{algorithm2e}

\begin{document}
%
\title{X-ray ghost tomography: denoising, dose fractionation and mask considerations}

%
%
%

\author{
Andrew~M.~Kingston$^{1,2,*}$,
Glenn~R.~Myers$^{1,2}$,
Daniele~Pelliccia$^{3}$,
Imants~D. Svalbe$^{4}$,
and~David~M.~Paganin$^{4}$
\thanks{$^1$ Dept. of Applied Mathematics,
Research School of Physics and Engineering,
The Australian National University,
Canberra, ACT 2601, Australia.
}
\thanks{$^2$ CTLab: National Laboratory for Micro Computed-Tomography,
Advanced Imaging Precinct,
The Australian National University,
Canberra, ACT 2601, Australia.
}
\thanks{$^3$ Instruments \& Data Tools Pty Ltd, Victoria 3178, Australia.
}
\thanks{$^4$ School of Physics and Astronomy, Monash University, Victoria 3800, Australia.
}
\thanks{$*$ Corresponding author, email: andrew.kingston@anu.edu.au}
\thanks{Manuscript received Month Day, 2018; revised Month Day, 2018.}
}

%
%

\markboth{Journal of Trans. on Computational Imaging,~Vol.~XX, No.~YY, Month~2018}%
{Kingston \MakeLowercase{\textit{et al}.}: X-ray ghost tomography}
%



\maketitle

\begin{abstract}

Ghost imaging has recently been successfully achieved in the X-ray regime; due to the penetrating power of X-rays this immediately opens up the possibility of X-ray ghost tomography. No research into this topic currently exists in the literature. Here we present adaptations of conventional tomography techniques to this new ghost imaging scheme. Several numerical implementations for tomography through X-ray ghost imaging are considered.  Specific attention is paid to schemes for denoising of the resulting tomographic reconstruction, issues related to dose fractionation, and considerations regarding the ensemble of illuminating masks used for ghost imaging.  Each theme is explored through a series of numerical simulations, and several suggestions offered for practical realisations of X-ray ghost tomography.

\end{abstract}


%
\IEEEpeerreviewmaketitle

\section{Introduction}
%
%
%
%
\IEEEPARstart{G}{host} imaging is an indirect imaging method, utilising intensity correlations, that originated in the domain of visible-light optics \cite{Klyshko1988, Belinskii1994, Pittman1995, Strekalov1995, Bromberg2009, katz2009compressive, Erkmen2010, Shapiro2012, Shirai2017}. The method has two key features.  Firstly, the only photons (or other imaging quanta such as X-rays \cite{Yu2016, pelliccia2016experimental, Schori2017, pelliccia2017practical, zhang2017table}, atoms \cite{Khakimov2017} {\em etc.}) that are registered by a position-sensitive detector are those that have never passed through the object.  A second feature is that imaging quanta that do pass through the object are never registered by a position sensitive detector.  Rather, quanta that have interacted with the object are measured with a large one-pixel detector that is commonly called a {\it bucket} in the ghost-imaging literature \cite{Moreau2017}.

\begin{figure}[!h]
\centering
\includegraphics[width=0.9\linewidth]{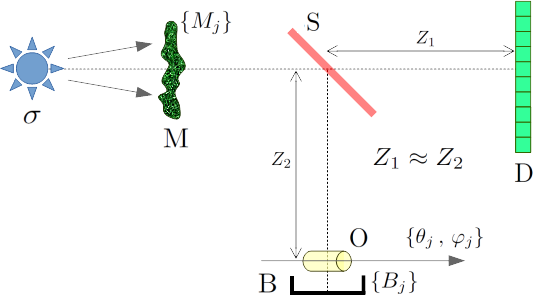}
\caption{
Generic experimental setup for X-ray ghost imaging and tomography.
\label{fig:2DXRGI}\label{fig:3DXRGI}}
\end{figure}

More precisely, a typical ghost-imaging scenario involves illuminating an unknown object with an ensemble of spatially highly-structured patterns, which may either be random speckle fields or engineered intensity distributions obtained using suitable coded apertures or spatial light modulators. The two-dimensional intensity distribution for each illuminating pattern is registered using the previously-mentioned position-sensitive detector. A beam-splitter is used to create a second copy of these illuminating patterns; this copy is typically much weaker in intensity than the parent beam \cite{zhang2017table}.  This second copy of the illuminating field is passed through an object of interest, downstream of which is placed the bucket detector that measures a single number (the {\it bucket signal}) for each illuminating pattern.  This bucket signal is proportional to the total number of imaging quanta transmitted by the object. A series of bucket signals, and corresponding intensity illumination maps, is recorded by varying the incident structured patterns. Refer to Fig. \ref{fig:2DXRGI} for a schematic of the experimental set-up.

Subsequently an image of the object is formed via the intensity {\em correlations} computed between the set of bucket measurements, and each pixel in the set of measured intensity illumination patterns. Ghost imaging is a quintessential example of computational imaging, since the computer forms an intrinsic component of imaging systems based on this principle. It is remarkable that such parallelized intensity--intensity correlations can form an image of the object, given that imaging quanta that have interacted with the object are never registered with a position-sensitive detector.

An overview of the origins of ghost imaging in both quantum and classical visible-light optics is beyond the scope of this paper.  We refer the reader to a recent review \cite{Moreau2017}.  Our focus is on the translation of ghost-imaging concepts to the X-ray domain, a relatively recent development for which there are only four currently-published papers as of this writing \cite{Yu2016,pelliccia2016experimental,Schori2017,zhang2017table}.  Notably, all published X-ray ghost-imaging reconstructions are in essence one-dimensional.  It is therefore timely to explore both two-dimensional and three-dimensional ghost imaging on account of the penetrating power of X-ray radiation, which enables both two-dimensional and three-dimensional images of optically opaque objects to be captured, and also on account of the {\AA}ngstr\"{o}m-scale wavelength of X-rays which imply an ultimate resolution limit on the scale of individual atoms.

Although computed tomography of optically opaque objects using X-rays has not been considered in literature, the concept of 3D ghost-imaging does indeed exist. Direct 3D tomographic imaging has been explored using ghost-imaging in an optical coherence tomography (OCT) context \cite{nasr2003demonstration}. Although limited to optically transparent objects, this quantum-OCT system has advantages in dealing with group-velocity dispersion (enabling imaging with a higher spectral bandwidth) and resolution given a Gaussian spectrum. Ghost topography, (or 3D imaging of a surface), using time-of-flight with a single-pixel camera was presented in \cite{sun2016singlePixel} in the context of remote sensing.

This work also has many similarities with single-pixel camera schemes ({\it e.g.}, \cite{duarte2008single}). Here, we are dealing with reflected visible light from a scene rather than a projected image, and instead of measuring each patterned illumination field in a primary beam, the structure of the illumination mask is controlled, and thus known {\it a-priori}. The success of such a single-pixel camera scheme, in combination with compressive sensing approaches to produce an image from fewer measurements is surely one of the keys to reducing dose in X-ray ghost imaging and tomography. This is achieved by leveraging the knowledge that the objective function (or object image) can be represented in some transform space with a sparse set of coefficients. A complementary technique for image recovery recently introduced in ghost-imaging is that of deep-learning \cite{lyu2017deep}; this is not explored here.

3D computed tomography using illumination masks (known as compressive tomography, {\it e.g.}, \cite{choi2009coded, brady2015compressive}) also has many similarities to the current work. Again, the illuminating masks are known {\it a-priori}, and here the {\it bucket} signal is actually a 2D image ({\it i.e.}, with a position sensitive detector). A principal difference is that in compressive tomography, reduced sampling (or measurements) applies to projected image space, while in ghost tomography the sampling can occur in any transform space of the projected images, (depending on the structure of the illumination patterns). However, the image/volume reconstruction techniques developed in the context of compressive tomography are generally applicable here.

We conclude this introduction with a brief summary of the remainder of the paper. The methods utilised in the work to simulate radiographic projection of a volume, illumination patterns, and the corresponding bucket signals are presented in Sec. \ref{sec:method}. Section~\ref{sec:ghost2d} treats X-ray ghost imaging in two spatial dimensions (2D).  The relative merits of four different approaches are compared, and some recommendations made for future experiments in 2D X-ray ghost imaging.  Section~\ref{sec:ghost3d-2step} considers a two-step approach to X-ray ghost tomography, in which one first separately reconstructs a series of two-dimensional projection X-ray ghost images and then combines the resulting series of 2D ghost-image projections into a 3D ghost tomogram.  For the ghost-imaging step, the same suite of algorithms is considered as was used in the preceding section; for the subsequent tomography step, two standard methods---filtered back-projection (FBP) \cite{kak1988principles}, and FBP followed by the simultaneous iterative reconstruction technique (SIRT) \cite{gilbert1972iterative}---are considered.  We also consider the use of compressed sensing (CS) techniques for exploiting sparsity to denoise the resulting reconstructions, as well as considering the question of dose fractionation.  Section~\ref{sec:ghost3d-direct} considers a single-step X-ray ghost tomography process that does not require the intermediate step of reconstructing 2D ghost projections.  An important tradeoff is identified, and dose fractionation again considered.  We pass to a consideration of masks in Sec.~\ref{sec:masks}, for both 2D and 3D X-ray ghost imaging, comparing the use of spatially random speckle masks with one particular class of coded mask.  Section~\ref{sec:discuss} discusses the possibilities for reduced dose in X-ray ghost imaging in comparison to direct X-ray imaging, makes some general remarks regarding random versus coded masks, and emphasises the desirability for different ensembles of mask to be used for different object orientations so as to avoid ring artifacts in X-ray ghost tomography.  Section~\ref{sec:FutureWork} outlines some possible avenues for future research.  We offer some concluding remarks in Sec.~\ref{sec:conclus}.

\section{Method: simulating radiographic projection data and illumination patterns}
\label{sec:method}

A generic schematic for X-ray ghost imaging in two spatial dimensions, ({\it i.e.}, X-ray transform of the volume or projection images of the object), is shown in Fig.~\ref{fig:2DXRGI} \cite{pelliccia2016experimental, Schori2017, pelliccia2017practical, zhang2017table}.  Here, an X-ray source $\sigma$ illuminates a thin transmissive mask $M$.  A beam-splitter $S$ creates two arms in the ghost-imaging setup.  The transmitted beam has its transverse spatial intensity structure registered by a position-sensitive detector $D$.  The reflected beam passes through a thin object $O$ before having the total intensity, transmitted by the object, recorded by a single-pixel detector $B$. This latter detector is usually termed a {\it bucket} detector, in a ghost imaging context \cite{shih2012physics}. Typically, the distance $Z_1$ from $S$ to $D$ will be made similar to the distance $Z_2$ from $S$ to the entrance surface of the object, since the spatial distribution of intensity transmitted by the mask will in general exhibit Fresnel diffraction, and it is important that the illuminating patterns measured at $D$ be equal to the intensity field illuminating the object, up to a multiplicative constant. 

\begin{figure}[!h]
\centering
\includegraphics[width=0.8\linewidth]{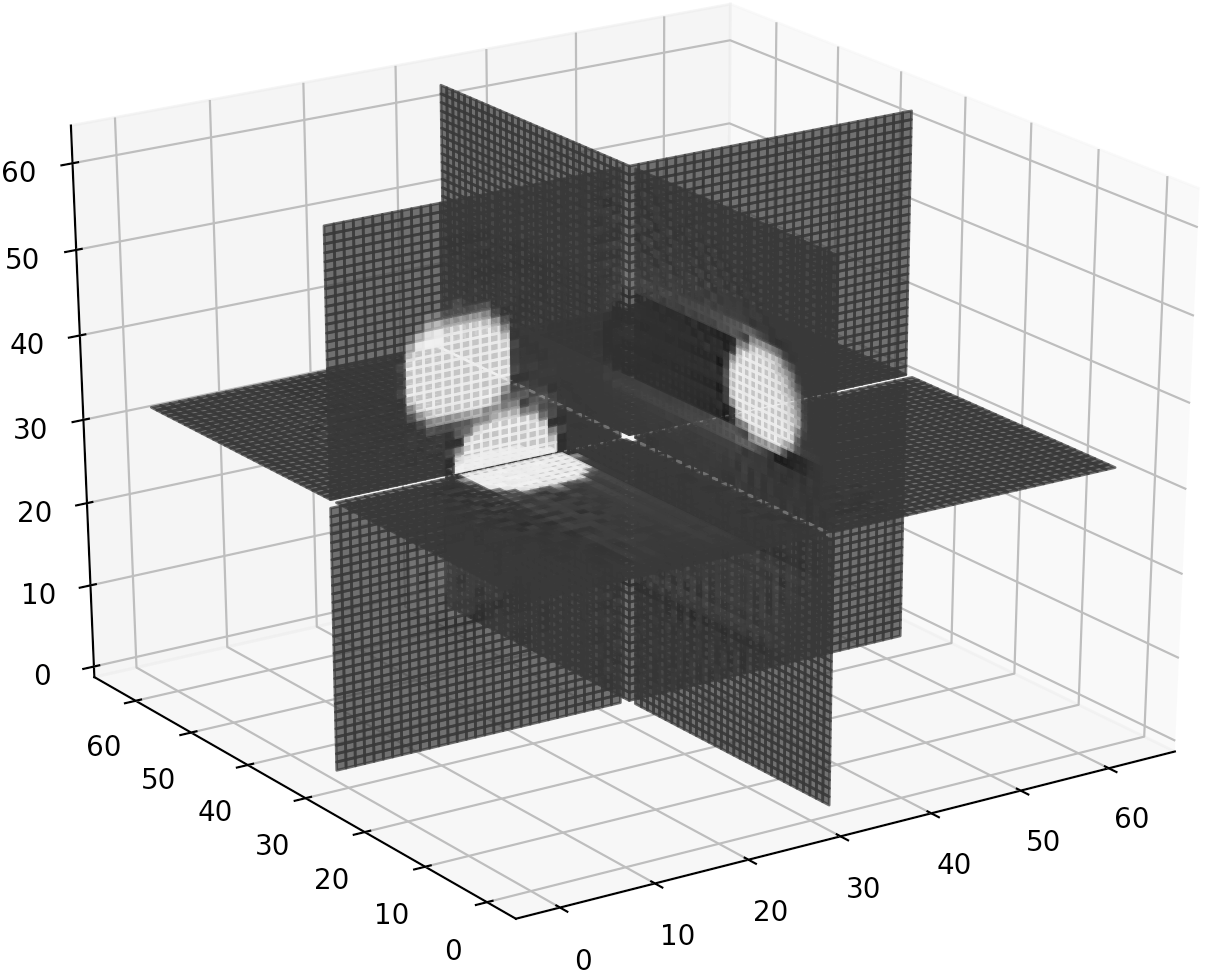}\\
\includegraphics[width=0.33\linewidth]{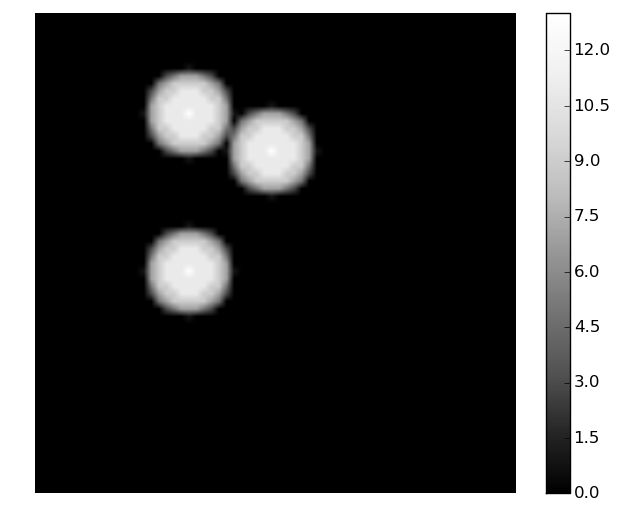}%
\includegraphics[width=0.33\linewidth]{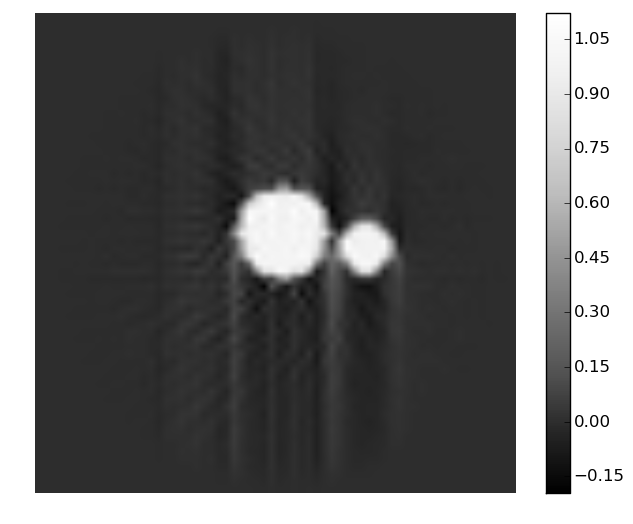}%
\includegraphics[width=0.33\linewidth]{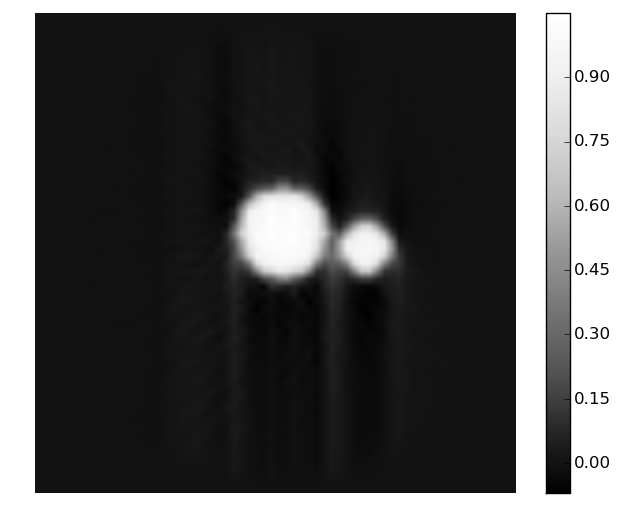}
\caption{
TOP: Three central orthogonal 2D slices through the $64 \times 64 \times 64$ voxel reconstructed tomogram determined using conventional direct-imaging X-ray tomography and filtered back-projection (FBP). BOTTOM: (L) The $64 \times 64$ pixel, $0^\circ$ simulated projection image of three identical spheres. (C-R) Slices through the 3D reconstructed volume at $r_3=18$; reconstructed by (C) FBP, and (R) 32 iterations of SIRT. All reconstruction are based on noise-free conventional imaging used for each projection.
\label{fig:orig_proj}
\label{fig:fbp_orig}
\label{fig:sirt_orig}}
\end{figure}

We turn now to simulated data for X-ray ghost tomography.  We are required to produce i) projected images of some 3D phantom volume, ii) a set of illumination masks, and iii) their corresponding bucket values by applying them to (i). The phantom considered here is comprised of three non-overlapping, identical spheres arranged in a 3D volume parameterised by Cartesian coordinates ${\bf r} = (r_1,r_2,r_3)$. The 3D X-ray linear attenuation coefficient $\mu({\bf r})$, is modelled using $\mathcal{N}^3$ cubic voxels with arbitrary dimension $\epsilon$ and $\mathcal{N}=64$. The spheres have a diameter of 12 voxels and an attenuation of 1.0 per voxel that can also be arbitrarily scaled. A depiction of this volume can be found in Fig. \ref{fig:fbp_orig}.

We assume a parallel X-ray beam geometry and thus use square pixels with the same arbitrary physical dimensions $\epsilon$ as the volume. The generic schematic for X-ray ghost imaging given in Fig.~\ref{fig:3DXRGI} applies here; note the object, $O$, is attached to a primary rotation stage providing azimuthal rotation and a secondary rotation stage providing polar rotation. The set of orientations is described in spherical polar coordinates by the polar angles $\theta_j$ and the azimuthal angles $\varphi_j$ corresponding to the $j^\mathrm{th}$ bucket measurement, $B_j$.

The transmitted intensity distribution in the contact plane for the $j$th bucket measurement $I'_{j} ({\bf x})$, is discretised on a grid $\mathcal{N}\times\mathcal{N}=64 \times 64$ pixels in size. The 2D Cartesian coordinates ${\bf x} = (x_1,x_2)$ parametrise the transverse position in the contact plane: the plane normal to the direction of X-ray propagation through the object, contacting the ``downstream'' side of the object. Since we are assuming a parallel beam geometry, rotation about a single axis provides sufficient information for exact reconstruction; the polar angle is considered to be constant, so that for the $j$th bucket measurement the object simply rotates by azimuthal angle $\varphi_j$, about the $x_2$-axis of the system. The $r_2$-axis of $\mu$ corresponds with the $x_2$-axis of the imaging system, and $\varphi_j$ describes the angle between the $r_1$-axis of $\mu$ and the $x_1$-axis of the system.

Projected attenuation, $A_\varphi ({\bf x})$, of the attenuation volume representing the object, $\mu({\bf r})$, at angle $\varphi$ is simulated by applying the X-ray projection transform $\mathcal{P}$ to each horizontal slice of the volume independently as follows:
\begin{eqnarray}
\label{equation:Xray_projection}
A_\varphi ({\bf x}) &=& \mathcal{P}_{\varphi}[\mu({\bf r})]\\
&=& \langle  \mu(r_1,x_2,r_3) | \delta (r_1 \cos \varphi - r_3 \sin \varphi - x_1)\rangle_{(r_1,r_3)},\nonumber
\end{eqnarray}
where $\langle|\rangle_{(r_1,r_3)}$ denotes the inner product spanned by Cartesian coordinate $(r_1, r_3)$.

This can be achieved numerically by rotating the volume about the system $x_2$-axis, or equivalently, the object $r_2$-axis, (perpendicular to the X-ray beam direction) by $\varphi$ and then summing rows along the beam axis. 
For each bucket measurement indexed by $j$, the transmission function of the object outlined above is
\begin{equation}
T_{\varphi_j}({\bf y}) \equiv \exp[-A_{\varphi_j} ({\bf x})],
\end{equation}
and the transmitted X-ray intensity in the contact plane $I'_{j} ({\bf x})$, is then found as
\begin{equation}
I'_{j} ({\bf x}) = I_j({\bf x}) T_{\varphi_j}({\bf y}),
\end{equation}
where $I_j({\bf x})$ is the incident X-ray intensity at position ${\bf x}$ in the contact plane. An example of a projected attenuation image at $\varphi = 0$ is given in Fig. \ref{fig:orig_proj}. To satisfy Nyquist angular-sampling requirements, $\pi \mathcal{N} / 2$ azimuthal angles are required \cite{crowther1970prsa}; we have thus generated 90 projected attenuation images with $\varphi$ distributed evenly over $\pi$ radians. 


Recall from Fig.~\ref{fig:2DXRGI} the thin transmissive mask $M$: for each mask in the $J$-member ensemble of masks $\{M_j\} \equiv \{M_j({\bf y})\}$, a bucket signal $B_j$ is measured, with corresponding intensity-distributions $I_j({\bf y})$ being measured by the position-sensitive detector $D$. Here, ${\bf y} = (y_1,y_2)$ denote Cartesian coordinates in planes orthogonal to the optic axis from source to detector.  Importantly, all X-rays registered by the position sensitive detector $D$ never pass through the object, while all X-rays in the arm containing the object are collected with a single-pixel bucket detector.  Moreover, the arm containing $D$ may be omitted altogether in a computational-ghost-imaging variant \cite{shapiro2008computational} where the masks $\{M_j({\bf y})\}$ are sufficiently well known that the the associated set of illuminating intensity patterns $\{I_j({\bf y})\}$ does not need to be measured because it may be {\em{calculated}} (e.g. using Fresnel diffraction theory) from knowledge of the set of mask or patterned images $\{M_j({\bf y})\}$.

Two classes of mask will be of particular interest in this study: (i) random masks whose transmitted intensity distribution is a spatially random speckle field. Each realisation has the same characteristic transverse length scale (speckle size) and root-mean-square (RMS) intensity at every point in the field of view of the mask. Every independent realisation $M$ of the mask is drawn from an ensemble $\{M_j({\bf y})\}$ containing $N$ independent masks \cite{pelliccia2017practical}; (ii) coded masks whose whose transmitted intensity distribution generates a linearly independent set of illuminating intensity maps, according to a deterministic algorithm such as is used to calculate uniformly redundant arrays \cite{fenimore}, so-called perfect arrays \cite{cavy2015construction}, {\em etc.}

The illuminating mask $M$ is again discretized using square pixels with arbitrary physical dimensions $\epsilon$. These illuminating $\mathcal{N}\times\mathcal{N}$-pixel masks $\{M_j({\bf y})\}$ constitute an ensemble of realisations of an $\mathcal{N}\times\mathcal{N}$ pseudo-random binary matrix with a mean of 0.5 \cite{RandomMatrixBook, CeddiaPaganin2018}. A spatially uniform flux of X-rays with intensity 1.0 is assumed such that the illuminating intensity patterns, $I_j({\bf y})$, generated by each mask are simply $I_j({\bf y}) = M_j({\bf y})$. Bucket signals are approximated via 
\begin{align}
\label{equation:GI_problem}
B_j \approx \langle I_j({\bf x}) | T_{\varphi_j}({\bf x}) \rangle_{\bf x}.
\end{align}
There are several important points to note here: (i) our simulations all assume zero noise in the photon detection process, {\it i.e.}, the effects of noise ({\em e.g.} photon shot noise, detector noise) and other detector imperfections ({\em e.g.} hot detector pixels, dead detector pixels, non-uniformities in detector gain, imperfections and losses of the beam-splitter) are not considered; (ii) The physical pixel/voxel size, $\epsilon$, does not need to be specified; (iii) the projected attenuation also has no scale. To simplify the processes of this numerical study we will assume that the phantom is {\it weakly absorbing}; this enables us to work directly with projected attenuation (as expanded below). An example phantom may be three poly(methyl-methacrylate), or PMMA, spheres of 2mm diameter imaged with 30keV X-rays. PMMA has a density of 1.18 g cm$^{-3}$ giving a linear attenuation coefficient of 0.072 cm$^{-1}$ or 93\% transmission. Transmission bucket measurements, $B_j$ can then be found directly from the projected attenuation bucket measurements, $B_j^A \approx \langle I_j({\bf x}) | A_{\varphi_j}({\bf x}) \rangle_{\bf x}$ by approximating the exponential term with the first order Maclaurin expansion as follows:
\begin{eqnarray*}
B_j & = & \langle I_j({\bf x}) | \exp[-A_{\varphi_j}({\bf x})]\rangle_{\bf x}\\
& \approx & \langle I_j({\bf x}) | 1-A_{\varphi_j}({\bf x})\rangle_{\bf x}\\
& = & \langle I_j({\bf x}) | 1\rangle_{\bf x} - B_j^A.
\end{eqnarray*}

This approximation greatly simplifies the following numerical study, since we can work directly with attenuation data, without compromising the validity and implications of the results. A significant difference does occur when tomographic reconstruction is performed directly from measured bucket values (Sec.~\ref{sec:ghost3d-direct}); the determination of error in projected attenuation from intensity bucket residuals will be presented for both cases (weak X-ray absorption and otherwise).

As a first benchmark for quality of tomographic reconstruction, the result of conventional (i.e. direct imaging) X-ray tomography using FBP is presented in Fig.~\ref{fig:fbp_orig}. Selected 2D slices in the $r_1-$, $r_2-$, and $r_3-$directions are presented. However, from this point on only the $r_3=18$ slice will be presented for comparison. Note that the mild streaks in the $r_3$-slice are due to a half-pixel offset introduced between the actual rotation axis, and the location of the axis used in the FBP reconstruction. This offset was kept in all remaining simulations. As a second benchmark, Fig.~\ref{fig:sirt_orig} shows the result of applying SIRT to the original projections for 2, 8, and 32 iterations. Regularisation can be achieved simply by reducing the number of iterations. In the case considered, 32 iterations are sufficient for a reasonable reconstruction.

\section{X-ray ghost imaging of radiographic projection images}
\label{sec:ghost2d}

We start this section by outlining the four different ghost-imaging (GI) approaches that are considered in the present paper. (i) The basic method for ghost image reconstruction, here termed the cross correlation (XC) method \cite{katz2009compressive} and written as operator $\mathcal{C}_{\varphi}$, approximates the intensity transmission function $T_{\varphi}({\bf x})$ of the object using
\begin{eqnarray}
\label{equation:XC_method}
T'_{\varphi}({\bf x}) &=& \mathcal{C}_{\varphi}(B_j)\\
&\equiv &\frac{1}{N}\sum_{j=1}^J (B_j - \overline{B}) I_j(x,y)\delta(\varphi - \varphi_j),
\end{eqnarray}
where, $\overline{B} = \frac{1}{J}\sum_{j=1}^J B_j$ is the average bucket reading, and $N$ is the number of buckets at angle $\varphi = \varphi_j$. (ii) A second approach can be formulated as follows: 
given a current estimate $T^k_{\varphi}({\bf x})$, (which may be obtained by XC), an improved estimate is obtained via the update scheme
\begin{equation}
T^{k+1}_{\varphi}({\bf x}) = T^{k}_{\varphi}({\bf x}) + \gamma\mathcal{C}_{\varphi}[B_j - \langle I_j({\bf x})|T^{k}_{\varphi_j}({\bf x})\rangle_{\bf x}].
\end{equation}
Identifying the ghost imaging operator as the adjoint of the cross correlation operator
\begin{equation}
\mathcal{C}^*_{\varphi_j}[T^{k}_{\varphi}({\bf x})] \equiv \langle I_j({\bf x})|T^{k}_{\varphi_j}({\bf x})\rangle_{\bf x}
\end{equation}
this can be written as
\begin{equation}
T^{k+1}_{\varphi}({\bf x}) = T^{k}_{\varphi}({\bf x}) + \gamma\mathcal{C}_{\varphi}\{B_j - \mathcal{C}^*_{\varphi_j}[T^{k}_{\varphi}({\bf x})]\},
\end{equation}
suggesting this is a gradient descent method where $\gamma$ is a Landweber relaxation factor. Note that $B_j - \mathcal{C}^*_{\varphi_j}[T^{k}_{\varphi}({\bf x})] \equiv B_j - B_j^k$ is the residual error from the currently estimated bucket values. We use $\gamma = \alpha / \sigma^2$ where $\sigma^2$ is the spatially averaged variance of $I_j({\bf x})$. Given complete data, we set $\alpha = 0.25$ and this reduces as the problem becomes more under-constrained.  This iterative cross-correlation (IXC) process is iterated until a suitable convergence criterion is achieved \cite{Landweber2018}. (iii) The third method uses conjugate-gradient \cite{Press2007} cross-correlation (CG-XC) to iteratively improve the reconstruction \cite{GI_with_CS_and_conjugate_gradients}. (iv) Compressed-sensing (CS) improvement to IXC is also considered (CS-CX), utilising various forms of sparsity constraint \cite{katz2009compressive, duarte2008single, brady2015compressive, katkovnik2012compressive, Qaisar2013}.  Three typical types of sparsity, relevant in the present context, are (a) image-space sparsity, where $T_{\varphi_j}({\bf x})$ is assumed to be negligible for most pixels, (b) gradient sparsity, for which $|\nabla_{\perp}T_{\varphi_j}({\bf x})|$ is negligible for most pixels, $\nabla_{\perp}$ being the gradient operator in the ${\bf x}$ plane, and (c) frequency-space sparsity, where $\mathcal{F}[T_{\varphi_j}({\bf x})]$ is negligible for most spatial frequencies, typically those with 
\begin{align}
\sqrt{k_{x_1}^2+k_{x_2}^2}\ge\kappa ,
\end{align}
where $\kappa$ is a cut-off spatial frequency, and $\mathcal{F_{\perp}}$ denotes Fourier transformation with respect to $x_1$ and $x_2$, with $k_{x_1}$ and $k_{x_2}$ being the corresponding spatial frequencies.  This CS process utilises the same step size $\gamma$ as defined previously.

\begin{figure}[!h]
\centering
\noindent
\includegraphics[width=0.4\linewidth]{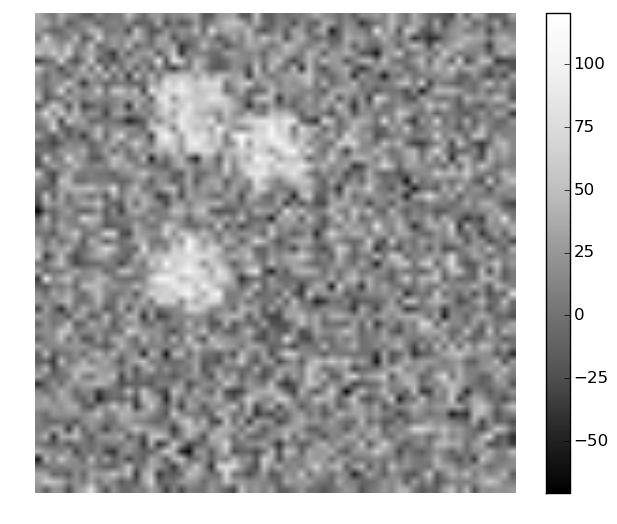}%
\includegraphics[width=0.4\linewidth]{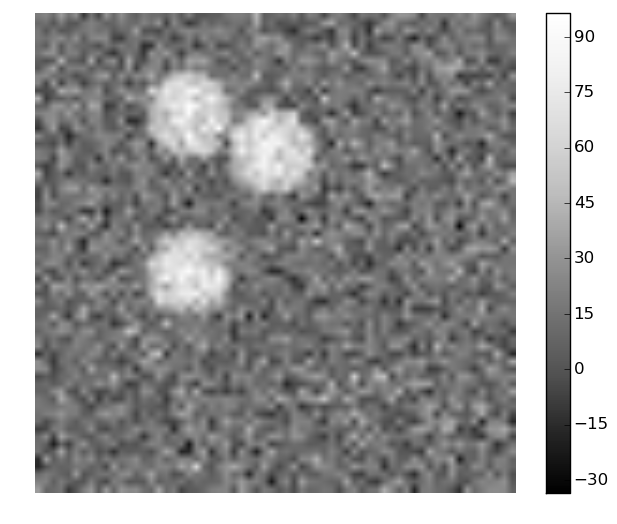}\\%
\includegraphics[width=0.4\linewidth]{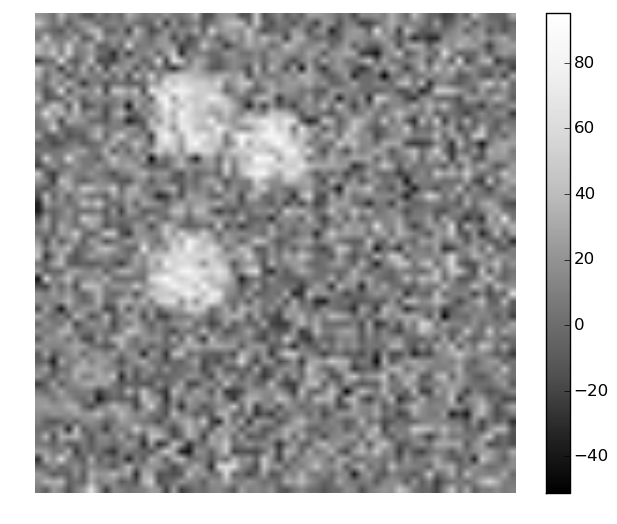}%
\includegraphics[width=0.4\linewidth]{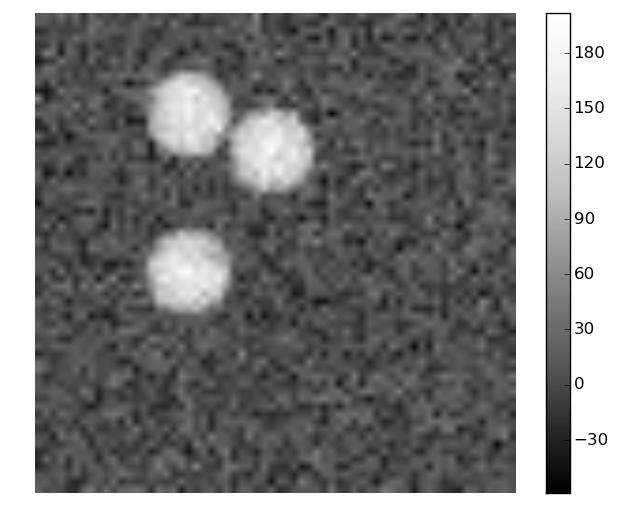}
\caption{
$64 \times 64$ pixel, $0^\circ$ X-ray ghost projection images recovered from (L--R) $J=1000, 4000$ measured bucket values generated with random binary masks with a mean value $\approx 0.5$. Recovery performed using (T--B) the XC method (MAD: 0.118, 0.0899), and 10 iterations of IXC (MAD: 0.101, 0.0682). Note: BL image used $\alpha = 0.025$.
\label{fig:ghost_proj_XC} \label{fig:ghost_proj_IXC}}
\end{figure}

Figure~\ref{fig:ghost_proj_XC} shows the simulations using the XC GI method to reconstruct the object transmission function in Fig.~\ref{fig:orig_proj}, for $J=1000, 2000, 3000, 4000$ masks respectively.  The normalised mean absolute deviation (MAD)---namely the mean absolute value of the difference between the reconstructed and input projection image (see Fig.~\ref{fig:orig_proj}), scaled to have a maximum value of 1.0, averaged over all pixels---is specified in the caption.  This error metric decreases as $J$ is increased.

The recovered images can be improved using the second method outlined above, namely IXC, as shown in Fig. \ref{fig:ghost_proj_XC}. For few measurements, convergence can be achieved by sufficiently reducing $\alpha$ to 0.025. 10 iterations of unaltered IXC for the $J=1000$ case increased the XC mean absolute deviation (MAD) from 0.118 to 0.185, while by reducing $\alpha$ to 0.025 ({\it i.e.}, scaled by 0.1), the XC MAD was reduced to 0.101 (again with 10 iterations). Still faster convergence can be achieved, for a moderate number of measurements, using the third means for 2D X-ray GI mentioned here, {\em i.e.} CG-XC.  However, for highly under-constrained data, stability becomes a problem, {\it e.g.} the MAD value after 10 iterations of CG-XC is 0.102 for the $J=1000$ case (worse than that for IXC).

One can improve on the IXC results for highly under-constrained or CS problems by leveraging knowledge of sparsity (in a suitable domain) to incorporate de-noising priors ({{\em cf.} Yao {\em et al.}, 2014 \cite{GI_denoising}). This is the fourth and final 2D GI method considered here. The three relevant sparsity constraints are image-space sparsity, gradient sparsity and frequency-space sparsity.  The results of applying these to the 1000-measurement image are shown in Fig. \ref{fig:ghost_proj_CS}. Here we have used 1000 iterations of IXC since for best results these priors are enforced very lightly and gradually, {\it e.g.}, with $\alpha = 0.01$. All sparsity assumptions yield significant improvement, with gradient sparsity seeming to be the most appropriate.

\begin{figure}[!h]
\centering
\noindent
\includegraphics[width=0.4\linewidth]{./crosscorr_recovery_1000.png}%
\includegraphics[width=0.4\linewidth]{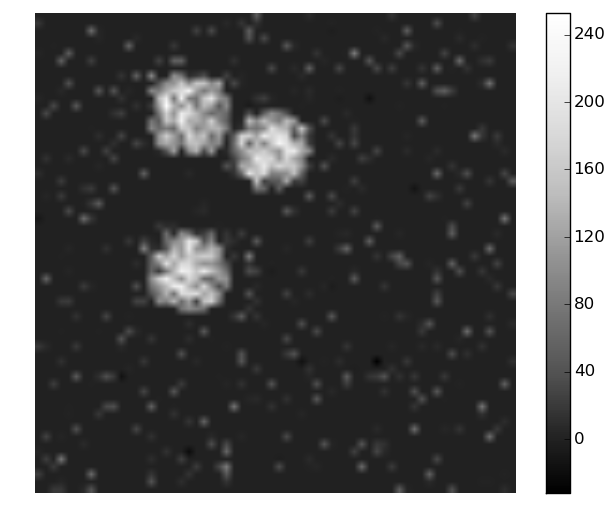}\\
\includegraphics[width=0.4\linewidth]{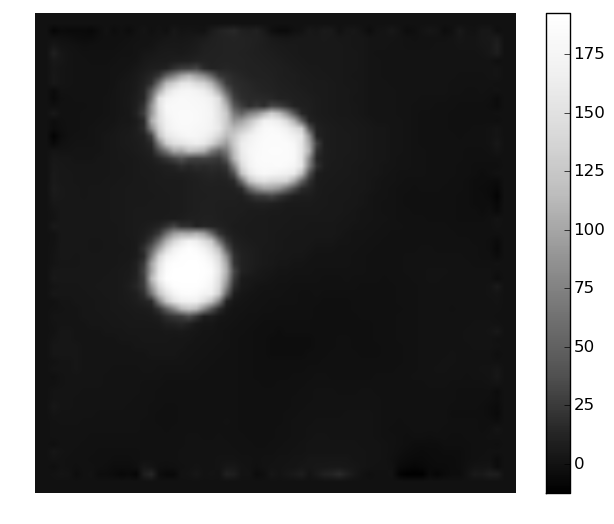}%
\includegraphics[width=0.4\linewidth]{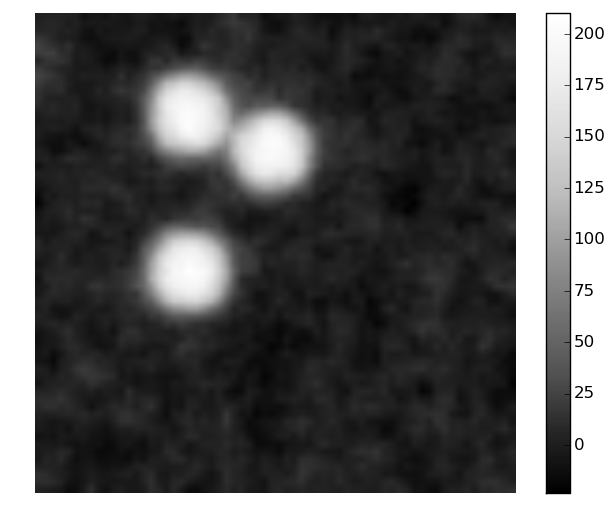}
\caption{
$64 \times 64$ pixel, $0^\circ$ X-ray ghost projection images recovered from 1000 measured bucket values generated with random binary masks with a mean value $\approx 0.5$. Recovery using 1000 iterations of IXC with sparsity assumptions: (clockwise) none, image-space sparsity, gradient sparsity, Fourier-space sparsity. (clockwise) MAD: 0.102, 0.0302, 0.0183, 0.0343.
\label{fig:ghost_proj_CS}}
\end{figure}

Based on these simulations, we draw the following conclusions regarding 2D X-ray ghost imaging: if IXC GI is used rather than XC GI, the method has a marginally improved MAD for a fixed number of masks, or a marginally reduced number of illuminating masks needed for a fixed required MAD. Faster convergence is possible if XC is combined with conjugate-gradient iterative refinement (CG-XC), however, this can become unstable for highly under-constrained cases.  For objects where sparsity assumptions may be employed, such as the object considered in the present study, compressed sensing approaches are the most favourable of the four X-ray GI approaches considered. 

\section{X-ray ghost tomography: two-step approach}
\label{sec:ghost3d-2step}

The two-step approach to 3D X-ray GI proceeds from bucket signals to the 3D reconstruction via the intermediate step of reconstructed 2D ghost-image projections. There are assumed to be $N$ images taken at each of $M$ object orientations $\varphi$, with $\{\varphi\}$ being an equally spaced subset of the 90 equally spaced azimuthal angles for which projection images are simulated.  For each orientation, the 2D ghost-image projection can be reconstructed using one of the four 2D GI methods introduced earlier (namely XC, IXC, CG-XC and CS-XC).  The second step is to use these retrieved X-ray ghost projections with two different standard tomographic approaches, namely (i) the analytic approach of filtered back-projection (FBP), or (ii) iterative refinement using SIRT.

\subsection{Effect of number of bucket measurements per projection}

We investigate the effect of altering the image quality of each X-ray ghost projection by changing the number of bucket value measurements $N$ per projection. The results from applying FBP to projections generated from 1000--4000 measurements are presented in Fig.~\ref{fig:fbp_ghost}. Similar results are obtained using FBP followed by 32 iterations of SIRT, as can be seen in Fig.~\ref{fig:sirt_ghost}. 

\begin{figure}[!h]
\centering
\noindent
\includegraphics[width=0.4\linewidth]{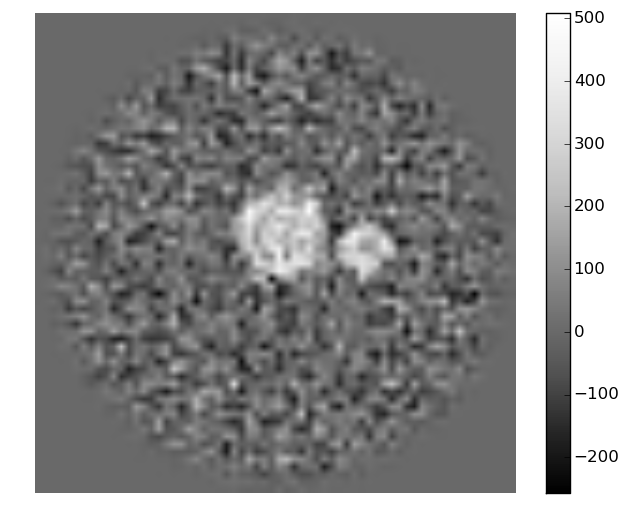}%
\includegraphics[width=0.4\linewidth]{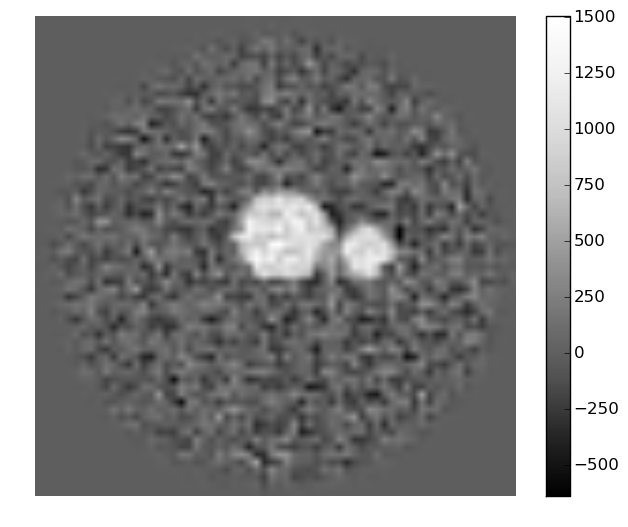}\\
\includegraphics[width=0.4\linewidth]{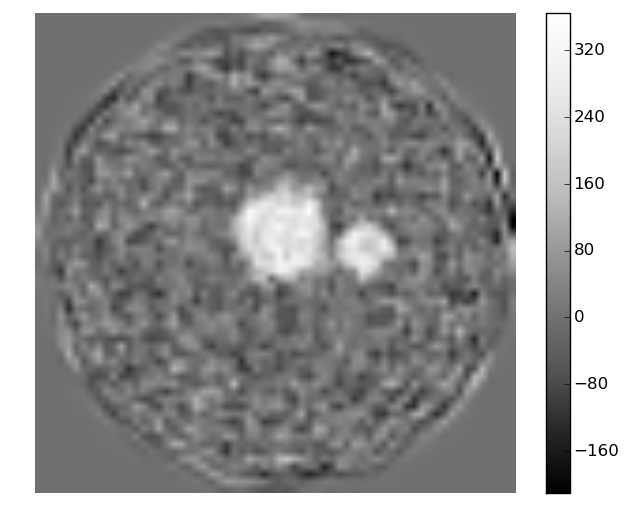}%
\includegraphics[width=0.4\linewidth]{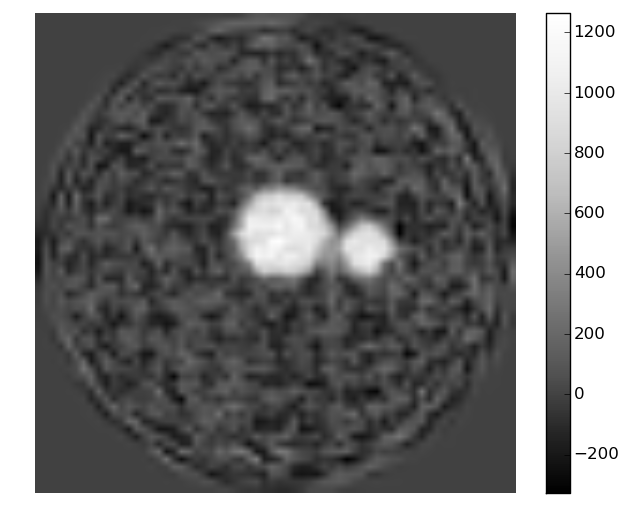}
\caption{
$r_3=18$ slice through the $64 \times 64 \times 64$ voxel 3D reconstructed volume. Generated from X-ray ghost projections recovered from (L--R) 1000, 4000 measured bucket values. Reconstruction performed using (T--B) FBP, and FBP followed by 32 iterations of SIRT.
\label{fig:fbp_ghost}\label{fig:sirt_ghost}}
\end{figure}

X-ray ghost tomogram quality improves with an increasing number of bucket measurements per projection, with SIRT refinement offering a marginal improvement compared to FBP alone.  

\subsection{Compressed sensing denoising techniques using sparsity}

The preceding 2D X-ray ghost simulations suggest that sparsity assumptions, where applicable, achieve rather better results than the marginal improvements obtained by IXC GI compared to XC GI.  The sparsity assumptions used here are: image-space sparsity via soft-thresholding, gradient sparsity via total variation (TV) minimisation \cite{GI_with_CS_and_conjugate_gradients}, and Fourier-space sparsity.  Each of these three sparsity assumptions is applied separately, followed by a fourth case where all three sparsity assumptions are applied. 

Henceforth we only compare the cases of $N=1000$ and $N=4000$ bucket measurements for each of the $M=90$ equally-spaced azimuthal angular orientations $\varphi$ of the sample.  A limit of 100 CS iterations has been adopted, since each such iteration is far more numerically costly in tomography when compared to 2D imaging.

Figure~\ref{fig:cs_ghost} shows the sparsity assumptions to be quite effective, particularly gradient and Fourier-space sparsity. There is little difference between the results obtained using $N=1000$ and $N=4000$ measurements per azimuthal object orientation, when using all three sparsity assumptions. This is consistent with the view that the dose-fractionation theorem \cite{hegerl_zn_1976} does not necessarily apply to CS CT.

\begin{figure}[!h]
\centering
\noindent
\includegraphics[width=0.4\linewidth]{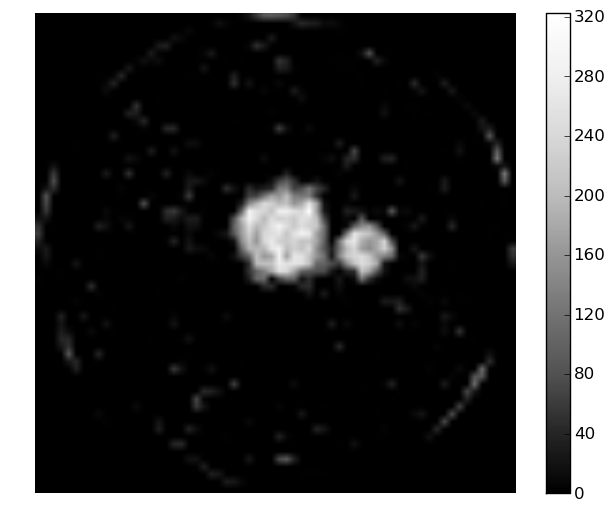}%
\includegraphics[width=0.4\linewidth]{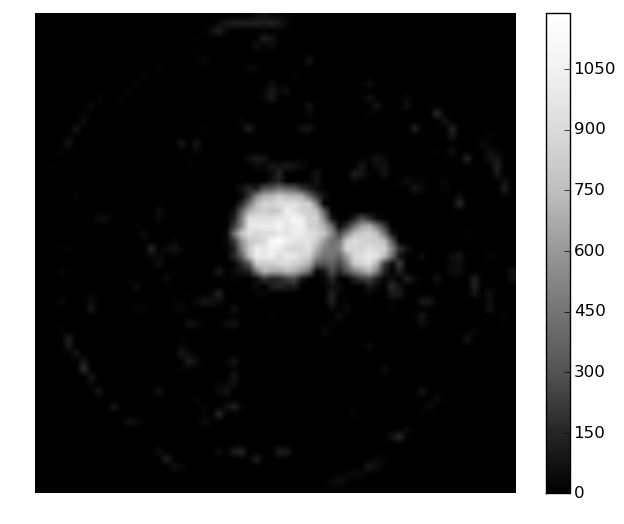}\\
\includegraphics[width=0.4\linewidth]{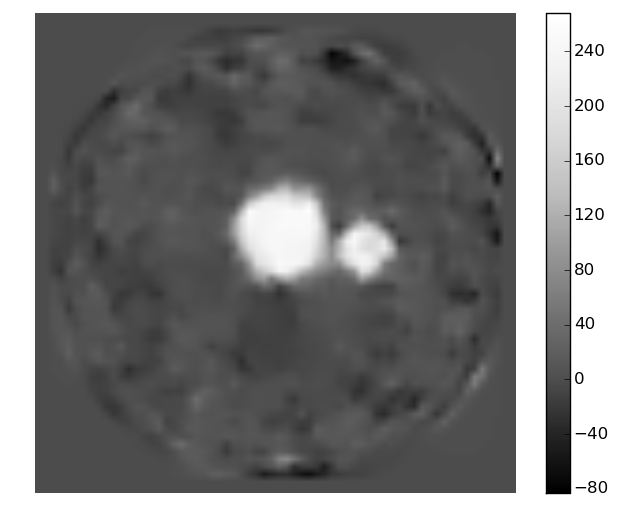}%
\includegraphics[width=0.4\linewidth]{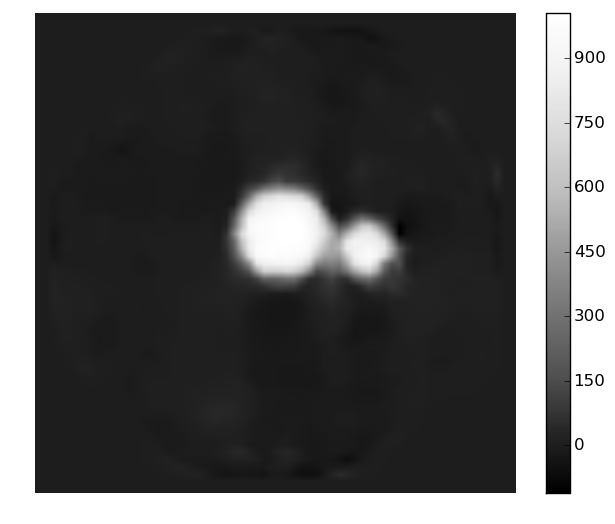}\\
\includegraphics[width=0.4\linewidth]{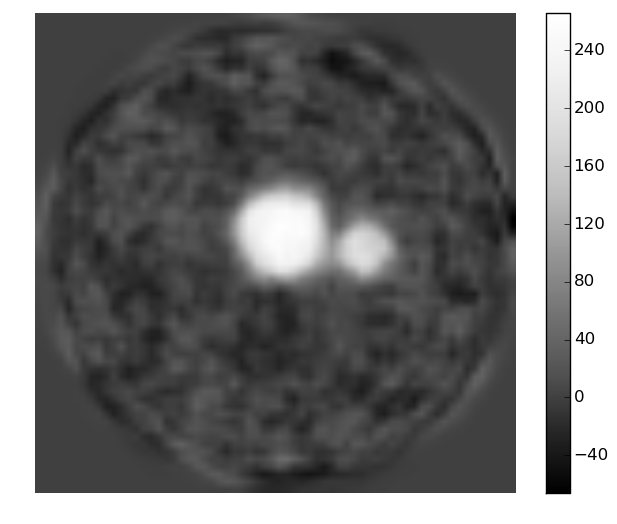}%
\includegraphics[width=0.4\linewidth]{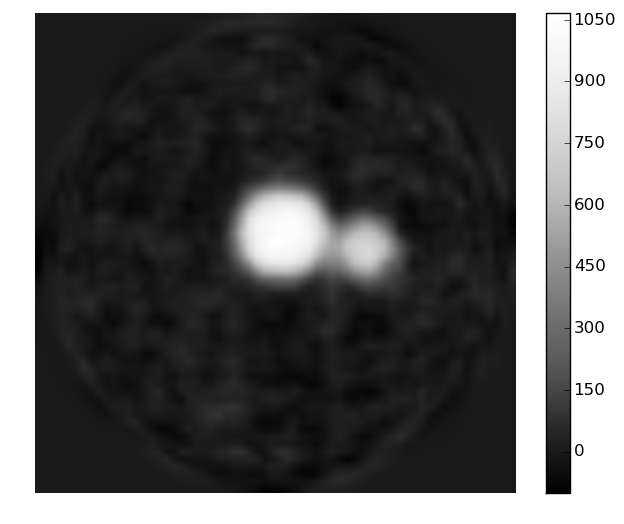}\\
\includegraphics[width=0.4\linewidth]{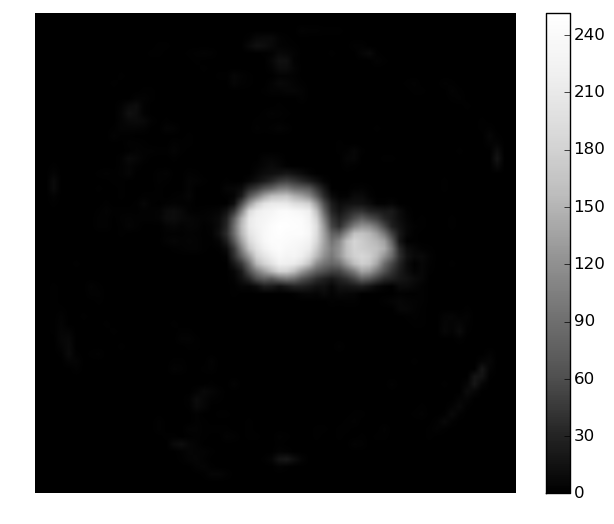}%
\includegraphics[width=0.4\linewidth]{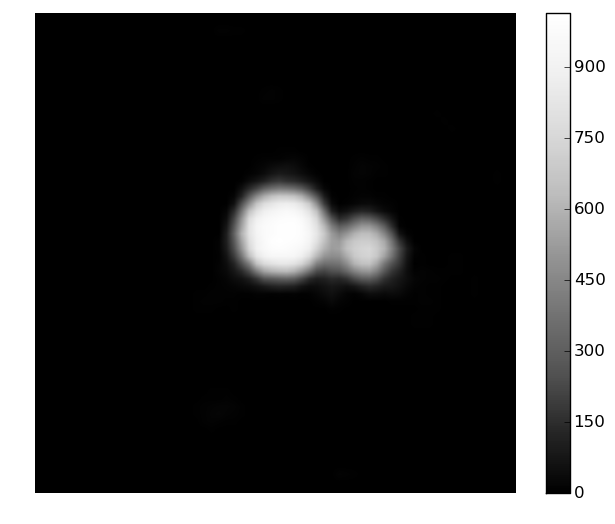}
\caption{
$r_3=18$ slice through the $64 \times 64 \times 64$ voxel 3D reconstructed volume. Generated from X-ray ghost projections recovered from (L--R) 1000 and 4000 measured bucket values per azimuthal angle $\varphi$. Reconstruction performed using 100 iterations of SIRT with (T--B) image-space sparsity via soft-thresholding, gradient sparsity via TV minimisation, Fourier-space sparsity, and all three sparsity assumptions.
\label{fig:cs_ghost}}
\end{figure}

\subsection{Dose fractionation}

Recall that in tomography, the total number of bucket measurements is $J=MN$, with $N$ illumination patterns being used at $M$ viewing angles. Assuming measurement time is constant, we can explore the effect of dose fractionation by keeping this total number, $J$, of bucket measurements constant, while varying the number $M$ of azimuthal orientations equally-spaced over $\pi$ radians.

Initially, we limit the numerical experiment to $J\approx 90,000$ bucket measurements, for $M=90$ projection images with $N=1000$ measurements each and $M=22$ projection images with $N=4000$ measurements each. The FBP tomographic slices in Fig. \ref{fig:fbp_m90000} again demonstrate the need for more sophisticated means of realising 3D X-ray GI in a practical manner, than can be provided by IXC and FBP alone, given the current experimental limitation that $J \lesssim 100,000$.

\begin{figure}[!h]
\centering
\noindent
\includegraphics[width=0.4\linewidth]{./recon_FBP_ghost_projections_1000__z18.png}%
\includegraphics[width=0.4\linewidth]{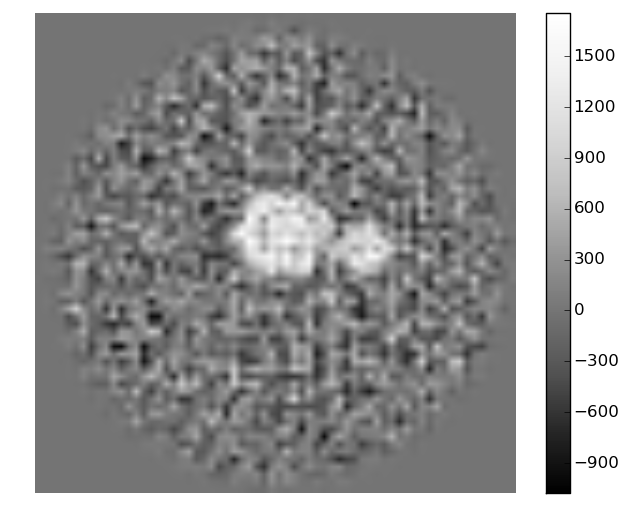}\\
\includegraphics[width=0.4\linewidth]{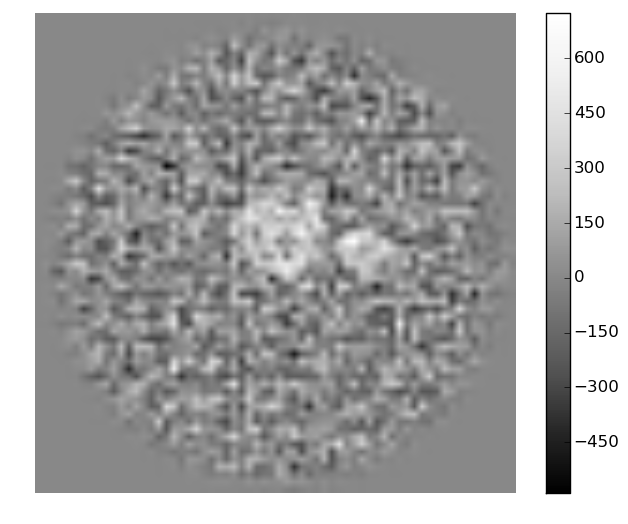}%
\includegraphics[width=0.4\linewidth]{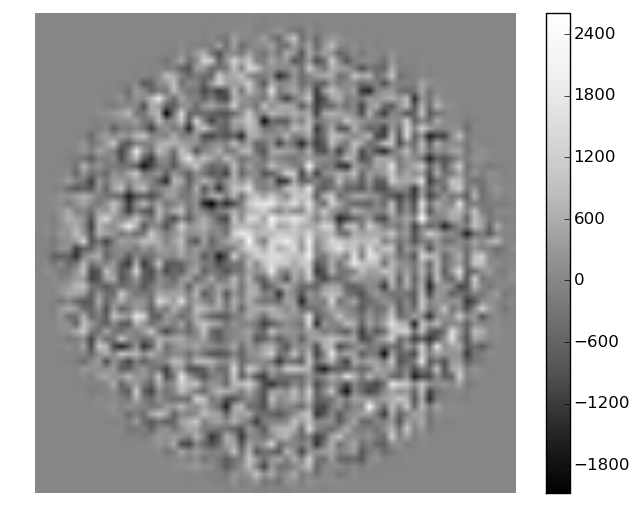}
\caption{
$r_3=18$ slice through the $64 \times 64 \times 64$ voxel 3D reconstructed volume. Generated from X-ray ghost projections recovered from (L--R) 1000, 4000 measured bucket values with a total of approx. (T--B) 90,000 and 30,000 measurements. Reconstruction performed using IXC and FBP.
\label{fig:fbp_m90000}\label{fig:fbp_m30000}}
\end{figure}

Next we further reduce the total number of bucket measurements to $J=30,000$, as $N=30$ projection images recovered from $M=1000$ measurements each, and $N=7$ projection images recovered from $M=4000$ measurements each. Again, for reference, the results from FBP are given in Fig. \ref{fig:fbp_m30000}. These reconstructed tomograms are too degraded for all but the most crude tomographic requirements: it is difficult to see the spheres against a noise-dominated background.

The results of utilising sparsity constraints, by subsequently applying 100 iterations of SIRT, are presented in Fig. \ref{fig:cs_m30000}. These reconstructions are comparable, although the case of $30 \times 1000$ measurements gives a slightly better resolving of the spheres than the case of $7 \times 4000$ measurements.  This similarity is to be expected, in light of the dose-fractionation theorem.

\begin{figure}[!h]
\centering
\noindent
\includegraphics[width=0.4\linewidth]{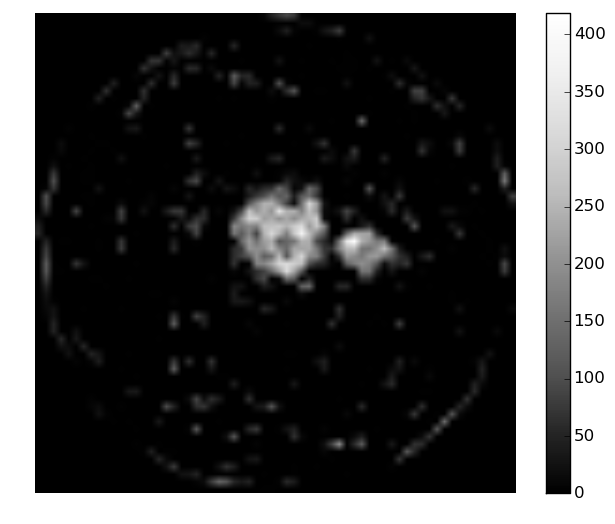}%
\includegraphics[width=0.4\linewidth]{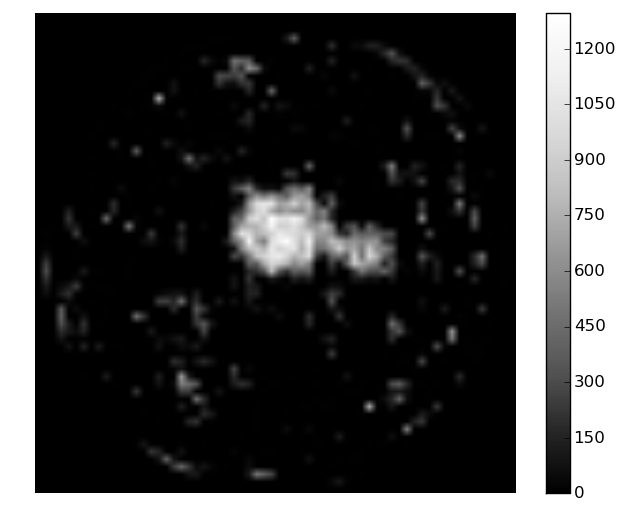}\\
\includegraphics[width=0.4\linewidth]{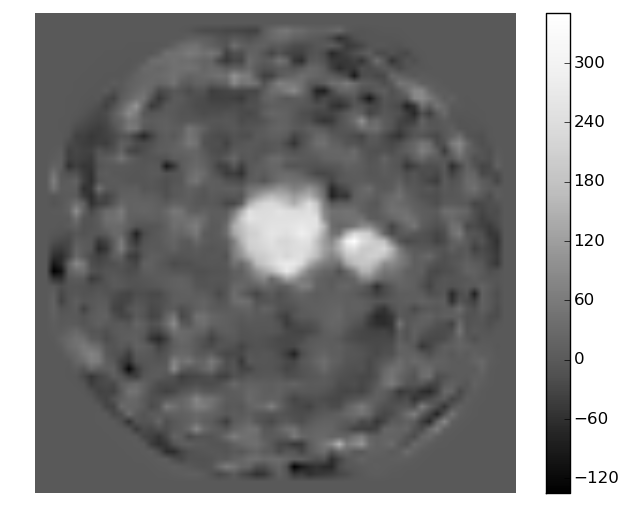}%
\includegraphics[width=0.4\linewidth]{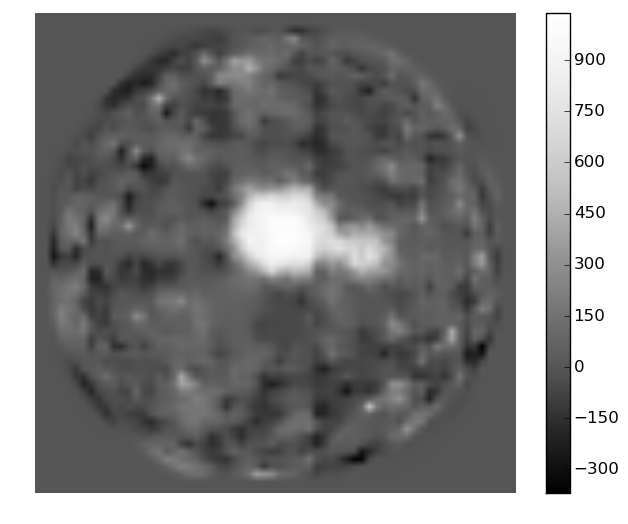}\\
\includegraphics[width=0.4\linewidth]{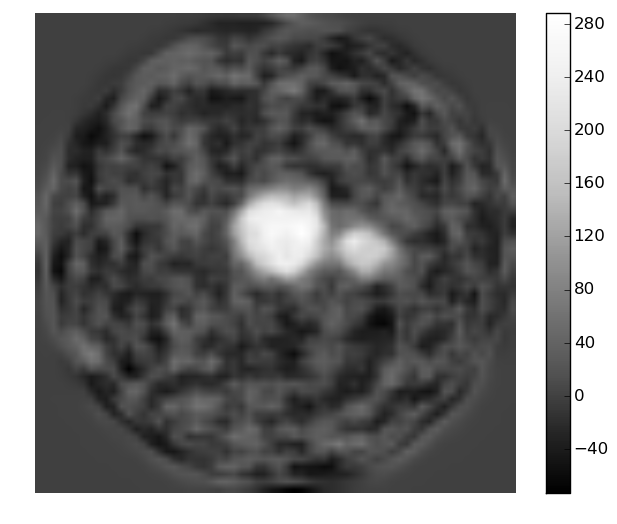}%
\includegraphics[width=0.4\linewidth]{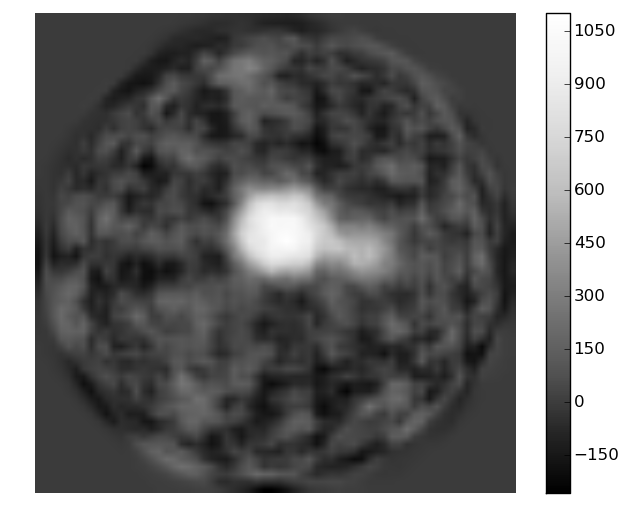}\\
\includegraphics[width=0.4\linewidth]{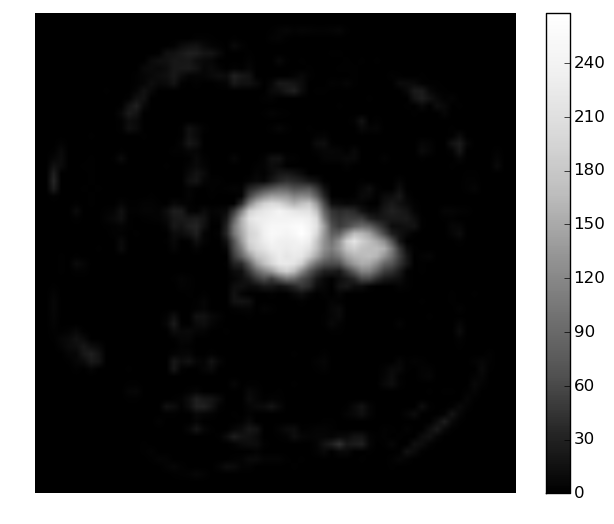}%
\includegraphics[width=0.4\linewidth]{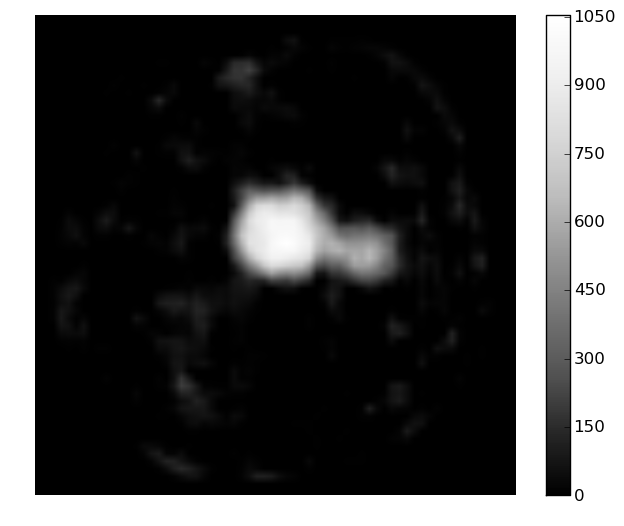}
\caption{
$r_3=18$ slice through the $64 \times 64 \times 64$ voxel 3D reconstructed volume. Generated from X-ray ghost projections recovered from (L--R) 1000 and 4000 measured bucket values with a total of approximately 30,000 measurements. Reconstruction performed using IXC and 100 iterations of SIRT with assumption of (T--B) image-space sparsity via soft-thresholding, gradient sparsity via TV minimisation, Fourier-space sparsity, and all three sparsity assumptions.
\label{fig:cs_m30000}}
\end{figure}

\section{X-ray ghost tomography: direct approach}
\label{sec:ghost3d-direct}

The preceding section considered a two-step approach to 3D X-ray GI by first determining a 2D GI reconstructed projection at each object orientation, with a subsequent tomographic step utilising the 2D GI projections.  A more direct approach, considered in the present section, proceeds directly from bucket signals to the 3D reconstruction, without the intermediate step of reconstructed 2D ghost-image projections.  

To motivate the direct approach to X-ray ghost tomography, we consider the question from an elementary perspective based on back-projection.  In direct-imaging filtered back-projection, each 2D projection of the object is back-projected along the line-of-sight into the reconstruction volume. This may be immediately generalised to give an elementary direct XC and SIRT form of X-ray ghost CT, by back-projecting
\begin{align}
\mathcal{C}(B_j) \equiv [B_j-\overline{B}] I_j({\bf x})
\end{align}
along the direction $(\theta_j,\varphi_j)$, ensemble averaging over all of the projections corresponding to all orientations $(\theta_j,\varphi_j)$ of the object (see Fig.\ref{fig:3DXRGI}; cf. Eq.~\ref{equation:XC_method}).  Note especially that the operator $\mathcal{C}$ is different from $\mathcal{C}_{\varphi}$ in the lack of a sum over buckets at angle $\varphi$. Recall that $I_j({\bf x})$ indicates the $j{\textrm th}$ illuminating speckle field, where the object is oriented at spherical polar angles $(\theta_j,\varphi_j)$, with $B_j$ being the corresponding bucket signal. Note we have used the weakly absorbing approximation here. This XC X-ray ghost CT formulation permits as few as one bucket signal per projection, {\it e.g.} if every successive orientation $(\theta_j,\varphi_j)$ is chosen randomly with a uniform distribution over the unit sphere.  This exemplifies the fact that it is not necessary to reconstruct 2D ghost projections prior to obtaining an X-ray ghost tomogram.

Of course, to avoid rotating the object every time a subsequent bucket measurement is taken, and also in light of simulations presented below, it will often be advantageous to take a series of bucket measurements for each object orientation. Even in this circumstance, the preceding argument makes it clear that it not necessary to proceed via an intermediate step of calculating 2D X-ray GI projections prior to undertaking the 3D ghost CT reconstruction.

In light of the preceding lessons regarding the importance of going beyond XC GI in an X-ray CT context, in the following simulations we have performed tomographic reconstruction directly from the measured bucket values, using a SIRT refinement to XC X-ray ghost CT. To perform simultaneous updates from all measured bucket values, the process is straight-forward, as outlined below. 

Recalling Eq.~\ref{equation:Xray_projection}, iterative tomographic reconstruction by gradient descent given a current estimate $\mu^k({\bf r})$ is then defined as:

\begin{equation}
\mu^{k+1}({\bf r}) = \mu^{k}({\bf r}) + \beta \mathcal{P}^* [A_\varphi ({\bf x}) - A^k_\varphi ({\bf x}) ], \label{eqn.SIRT}
\end{equation}
where $\beta$ is again a Landweber relaxation factor (typically set to $J^{-1}$). Here, $\mathcal{P}^*$ is the back-projection operator (the adjoint of the projection operator). The determination of the residual in projected attenuation, $A_\varphi ({\bf x}) - A^k_\varphi ({\bf x})$, depends on the relevant assumptions and simplifications. In our simulations, we have utilised the weakly absorbing approximation, {\it i.e.}, $T_\varphi({\bf x}) = \exp[-A_\varphi({\bf x})] \approx 1 - A_\varphi({\bf x})$, yielding:
\begin{eqnarray*}
A_\varphi ({\bf x}) - A^k_\varphi ({\bf x}) & = & T^k_\varphi ({\bf x}) - T_\varphi ({\bf x})\\
& = & \mathcal{C}_{\varphi}(B^k_j - B_j)\\
& = & \mathcal{C}_{\varphi}[\langle I_j({\bf x})|T^k_{\varphi_j} ({\bf x})\rangle_{\bf x} - B_j]\\
& = & \mathcal{C}_{\varphi}[\langle I_j({\bf x})|1-\mathcal{P}_{\varphi_j}\mu ({\bf r})\rangle_{\bf x} - B_j]\\
& = & \mathcal{C}_{\varphi}\{\mathcal{C}_{\varphi_j}^*[1-\mathcal{P}_{\varphi}\mu ({\bf r})] - B_j\}
\end{eqnarray*}
Noting that the backprojection operator $\mathcal{P}^*$ contains a normalised summation over viewing angles and that the Radon transform is linear, we can re-write equation \ref{eqn.SIRT} in the form discussed at the start of this section:
\begin{equation}
\mu^{k+1}({\bf r}) = \mu^{k}({\bf r}) + \beta \mathcal{P}^* \mathcal{C}\{\mathcal{C}_{\varphi_j}^*[1-\mathcal{P}_{\varphi}\mu ({\bf r})] - B_j\}.
\end{equation}
Without the weakly-attenuating approximation, the required correction in projected attenuation is determined as follows:
\begin{eqnarray*}
A_\varphi ({\bf x}) - A^k_\varphi ({\bf x}) 
& = & \log (\mathcal{C}_{\varphi}B_j) - \log \{\mathcal{C}_{\varphi}\mathcal{C}_{\varphi_j}^*\exp[\mathcal{P}_{\varphi}\mu ({\bf r})]\}.
\end{eqnarray*}

\subsection{Comparison with IXC followed by SIRT}

The result from applying 10 iterations of the above algorithm can be seen in the bottom row of Fig.~\ref{fig:sirt_bucket_10-10}. The normalised root mean squared error (RMSE) in the estimated bucket values after these 10 iterations is 8.44, 8.67, 8.74, and 8.78 ($\times 10^{-3}$) for (L--R) 1000, 2000, 3000, and 4000 measured bucket values per viewing angle. Normalisation in this case means scaled by expected bucket signal ($0.5 \times 3 \times \frac{4}{3} \pi 6^3$). This can be compared with the result from performing 10 iterations of cross-correlation (IXC) followed by 10 iterations of standard SIRT (top row of Fig. \ref{fig:sirt_bucket_10-10}); the RMSE in this case is 12.6, 9.46, 8.13, and 7.22 ($\times 10^{-3}$).

\begin{figure}[!h]
\centering
\noindent
\includegraphics[width=0.33\linewidth]{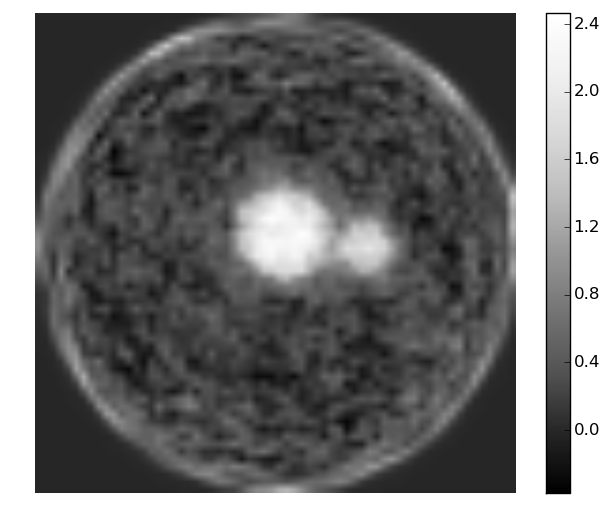}%
\includegraphics[width=0.33\linewidth]{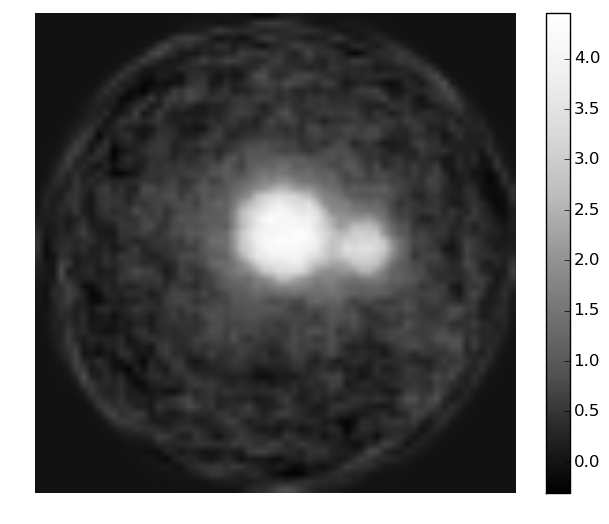}%
\includegraphics[width=0.33\linewidth]{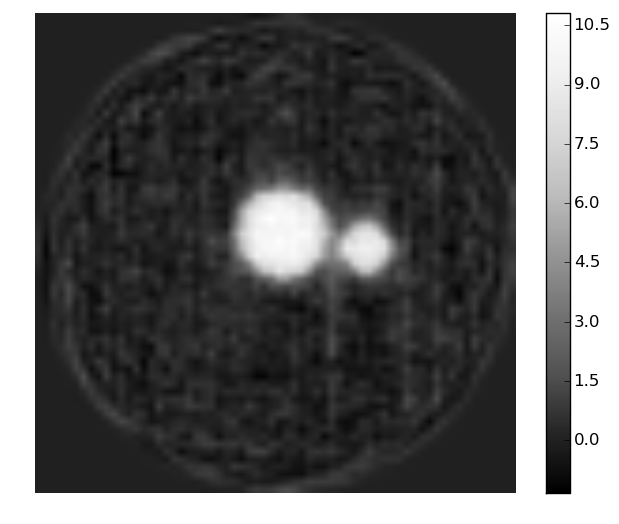}
\caption{
$r_3=18$ slice through the $64 \times 64 \times 64$ voxel 3D reconstructed X-ray ghost volume, generated directly from measured bucket values with 90 equally spaced azimuthal viewing angles. Reconstruction performed using 10 IXC with 10 iterations of SIRT, and 10 iterations of the algorithm described above with 1000 bucket values per viewing angle. (R) The improvement from 100 iterations of SIRT.
\label{fig:sirt_bucket_10-10}\label{fig:sirt_bucket}}
\end{figure}

This clearly shows the power of tomography directly from the measured bucket values, with regard to the suppression of noise. The bottom row of Fig. \ref{fig:sirt_bucket_10-10} is clearly less noisy and is reflected in the much lower MSE for the noisy cases (left). The higher MSE in the more well constrained problems is due to the above algorithm converging more slowly; high-frequency terms are still converging.

Better results can be achieved in fewer SIRT iterations by performing multiple iterations of cross-correlation in step 4 of the above algorithm. However, this step is more computationally expensive than projection/back-projection, {\em i.e.}, O($N^5$) compared with O($N^4$) operations, and overall iteration time is longer.  This is a typical trade-off between reconstruction quality and associated computational expense. Of course, results also improve with more iterations of SIRT as shown on the right of Fig. \ref{fig:sirt_bucket} where 100 iterations are performed.

\subsection{Dose fractionation}

Here we limit the total number of bucket measurements to $J=30,000$. Similar to the previous dose-fractionation section, $N=30$ and $N=10$ viewing angles are considered, with $M=1000$ and $M=3000$ measurements each respectively. The result from 100 SIRT iterations is presented in the second and third panels of Fig.~\ref{fig:sirt_bucket_m30000}. The power of this technique is again evident, as these results are produced without using any CS assumptions. This may be compared with the FBP results in Fig.~\ref{fig:fbp_m30000} and the CS results in Fig.~\ref{fig:cs_m30000}.

\begin{figure}[!h]
\centering
\noindent
\includegraphics[width=0.4\linewidth]{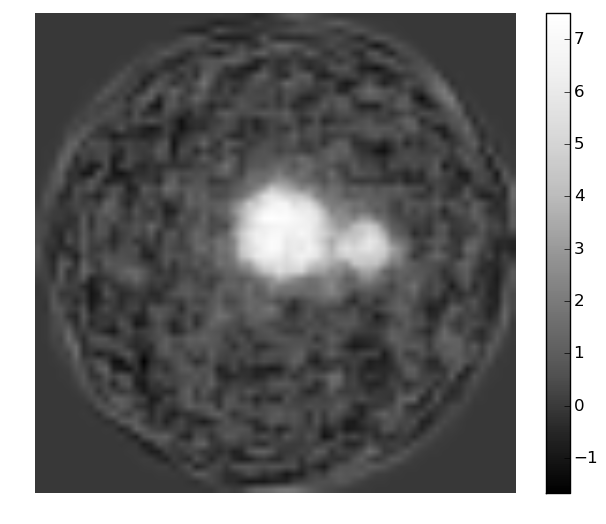}%
\includegraphics[width=0.4\linewidth]{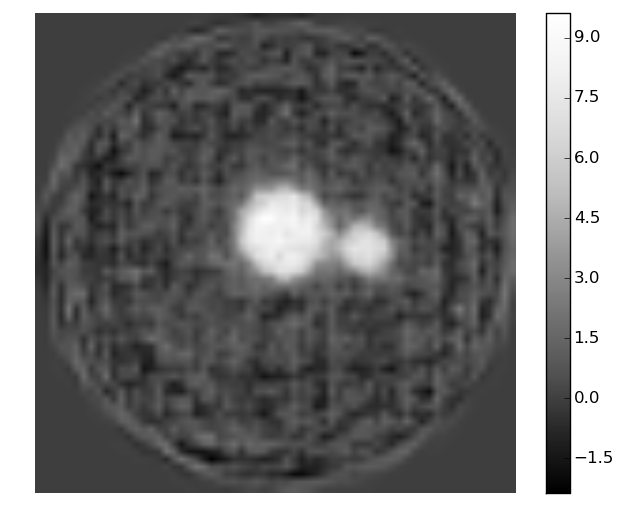}\\
\includegraphics[width=0.4\linewidth]{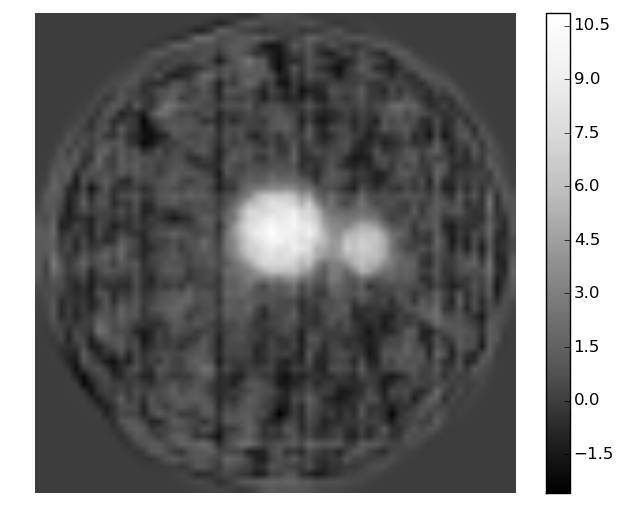}%
\includegraphics[width=0.4\linewidth]{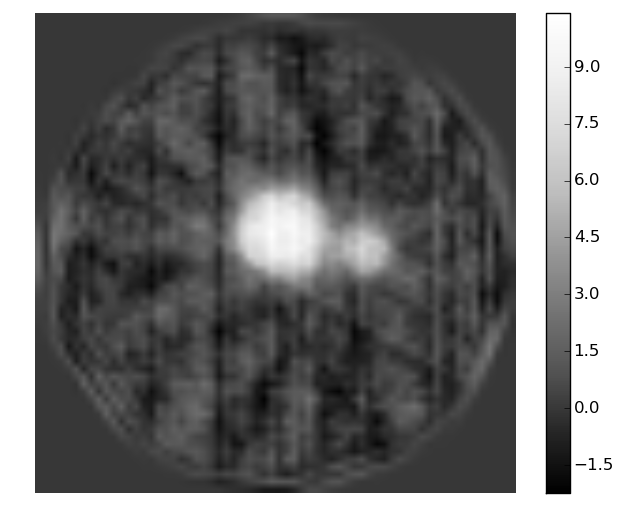}
\caption{
$r_3=18$ slice through the $64 \times 64 \times 64$ voxel 3D reconstructed X-ray ghost volume, generated directly from a total of 30,000 measured bucket values. Reconstruction performed using 100 iterations of SIRT described above with (clockwise) 333, 1000, 3000, and 4000 bucket values per viewing angle.
\label{fig:sirt_bucket_m30000}}
\end{figure}

Also included is data for 90 and 7 viewing angles (with 333 and 4000 bucket measurements each, respectively) on the far left and far right panels of Fig.~\ref{fig:sirt_bucket_m30000}.  These additional simulations explore the limits of dose fractionation. If there are too few bucket values per viewing angle then noise starts to dominate in each recovered projection image, resulting in a noisy and coarsened reconstructed volume (far left panel of Fig.~\ref{fig:sirt_bucket_m30000}).  Conversely, too few viewing angles results in streaking artifacts (far right panel of Fig. \ref{fig:sirt_bucket_m30000}). The best results in this study, limiting the total number of bucket measurements to 30,000, correspond to 30 equally-spaced azimuthal views, each of which have 1000 bucket measurements.

We emphasize the key message of noise suppression, for X-ray ghost tomography employing IXC followed by SIRT.  There is an evident trade-off in quality of reconstruction versus reconstruction time, relating to the number of iterations of IXC followed by the number of iterations of SIRT.

\section{Mask considerations in 2D and 3D X-ray ghost imaging}
\label{sec:masks}

Thus far we have utilised spatially random masks in all simulations.  These correspond to illuminating intensity distributions given by an ensemble of $J=MN$ random $\mathcal{N}\times\mathcal{N}$ matrices, each element of which is an independent deviate drawn from the same uniform probability distribution over the interval from zero to unity.  Generalisations of this include different probability distributions such as normal or Poisson processes, and/or introducing coupling between adjacent pixels via smearing each member of the random-matrix basis with a suitable discrete convolution kernel \cite{CeddiaPaganin2018}.  These generalisations, while interesting, will not be considered here.  

It is important, in the context of an investigation of the functional form adopted for the illuminating masks, to note that the standard XC ghost formula in Eq.~\ref{equation:XC_method} considers the set of linearly independent illumination patterns to form a complete or near-complete (indeed, over-complete if $J > \mathcal{N}^2$) mathematical basis for $\mathcal{N}\times\mathcal{N}$-pixel images.  Moreoever, there is an implicit assumption, in this formula, that the set of background-subtracted illumination patterns $\{I_j({\bf x})-\overline{I}\}$ approximate an orthogonal basis \cite{pelliccia2017practical,CeddiaPaganin2018}; here, $\overline{I}$ denotes the spatial average of the intensity for each illuminating pattern, with this spatial average assumed to be approximately the same for all illuminating patterns. This observation leads one naturally to consider non-random orthogonal (or near-orthogonal) masks \cite{GureyevGhost2018}.  This is the topic of the present section.


There are a wide variety of non-random masks that could be studied.  Indeed, there is an infinite multiplicity of orthogonal masks that could be devised for reconstructing pixelated arrays of a specified size \cite{GureyevGhost2018}.  Moreover, since such masks are deterministic, if their associated transmission functions $I_j({\bf x})$ are sufficiently well known then the detector $D$ in Fig.~\ref{fig:2DXRGI} may be eliminated altogether, thus transitioning 2D and 3D X-ray ghost imaging to 2D and 3D X-ray computational imaging.

In the simulations presented below, we compare the previously considered random masks, with two classes of non-random mask (coded aperture) that have near-perfect autocorrelation.  We then examine their relative performance in (2D) X-ray ghost imaging and (3D) X-ray ghost tomography.  We examine an issue that arises, pertaining to ring artifacts, and consider a means to avoid such artifacts in practice. 

\subsection{Coded-aperture generation}

In practical applications of X-ray ghost imaging, we do not use a random set of masks each of which are independently generated for each bucket reading, but rather we scan a single mask over the sample \cite{Schori2017,pelliccia2017practical,zhang2017table}.  This single mask, be it random or deterministic, is displaced in between bucket readings by an integer number of pixels in either of two perpendicular transverse directions. We assume that the masks are periodic in $y_1$/$y_2$, with respective periods of $\mathcal{N}\Delta_{y_1}$ and $\mathcal{N}\Delta_{y_2}$, hence for an $\mathcal{N} \times \mathcal{N}$ mask there are $\mathcal{N}^2$ possible unique bucket measurements that can be made.

A mask that has perfect auto-correlation \cite{cavy2015construction}, {\em i.e.}, zero at all values except the origin, produces an orthogonal basis in this case. There are many ways to produce such masks.  We investigate two methods: (i) modified uniformly redundant arrays (MURA) constructed using quadratic residues \cite{Gottesman1989newFamily}; (ii) a method based on the finite Radon transform (FRT) \cite{cavy2015construction}. We will compare their performance against a random mask, for both 2D and 3D X-ray ghost imaging, in the following subsections. These masks are constructed for $\mathcal{N}=59$, which gives 3481 elements. This is an odd number, therefore, we cannot have $\{-1,1\}$ cancelling to precisely give zeros in the auto-correlation, hence we will have {\em{near}} perfect auto-correlations, that are equal to $\pm 1$ away from the origin.

In fact, owing to the restriction that intensity maps must be non-negative and intensity transmission masks are restricted to values between zero and unity, we will not have zero cross-correlations at all, but rather $\mu \sigma^2 N^2$ where $\mu$ = mean, $\sigma^2$ = variance, and $N^2$ = mask size. They can be orthogonal if the mean is subtracted, this being an important requirement in the context of ghost imaging \cite{pelliccia2016experimental,pelliccia2017practical,CeddiaPaganin2018,GureyevGhost2018}. 

The masks investigated here, along with their auto-correlations, are given in the left and middle columns of Fig.~\ref{fig:coded_apertures}, respectively. All auto-correlations appear similar with a peak value of approximately 3481/2. These auto-correlations are proportional to the point spread function (PSF) of each set of masks \cite{pelliccia2017practical,GureyevGhost2018, ferri2010}.  These auto-correlations are not perfect in that they are not precisely proportional to a spatial Kronecker delta at the origin. However, the deviation from a spatial Kronecker delta is mild.  Indeed, when we replace the peak of each auto-correlation with its mean value, as shown in the right column of Fig.~\ref{fig:coded_apertures}, we observe that the random mask off-origin correlations have a range of 100 (which is greater than 10\% of the mean) while the coded-apertures only have a range of 1.

\begin{figure}[!h]
\centering
\noindent
\includegraphics[width=0.33\linewidth]{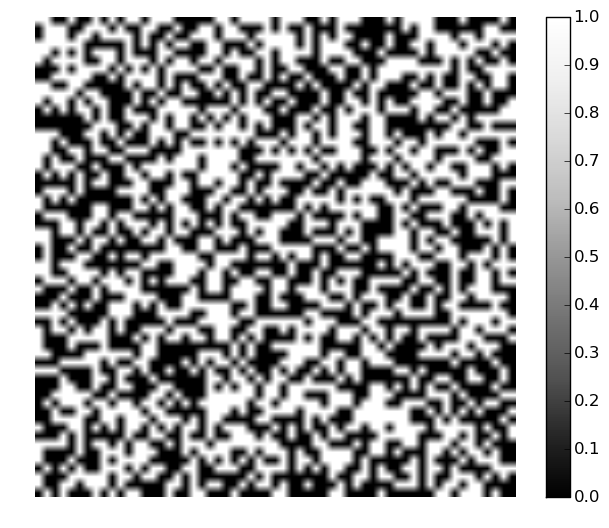}%
\includegraphics[width=0.33\linewidth]{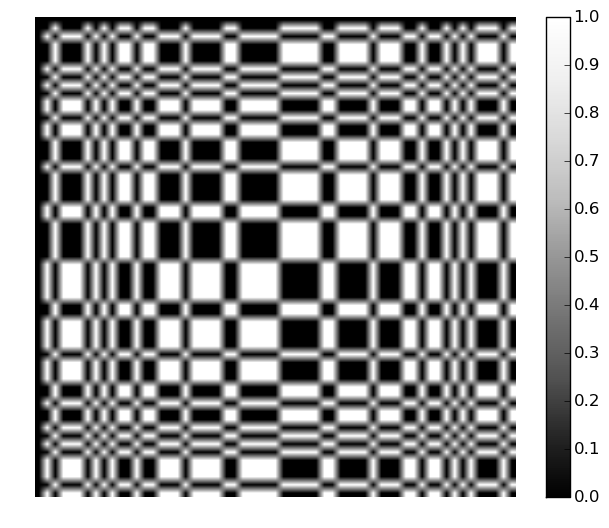}%
\includegraphics[width=0.33\linewidth]{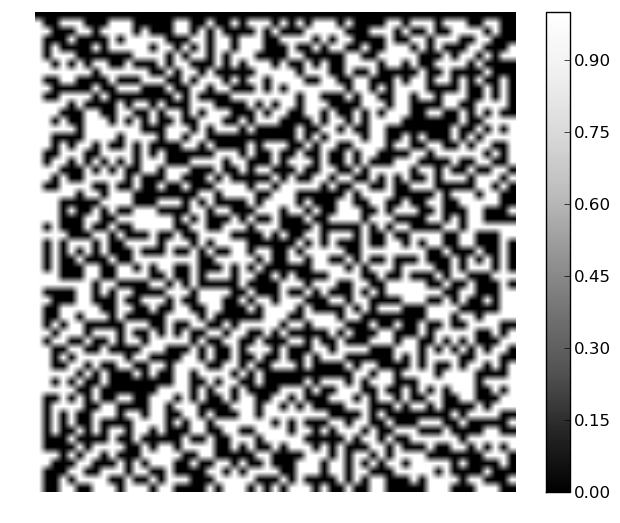}
\caption{
Examples of the types of mask explored here: (L-R) random mask, quadratic residue mask, finite Radon transform based mask.
\label{fig:coded_apertures}}
\end{figure}

\subsection{Performance of coded and random masks in X-ray ghost imaging}

\subsubsection{Radiographic performance}

The results of XC GI are presented in Fig. \ref{fig:ghost_imaging_coded_apertures}. The respective rows (T--B) of this figure use random masks, quadratic-residue coded masks and FRT-based coded masks.  Respective columns (L--R) use XC applied to 3481, 2610, 1740 and 870 bucket measurements.  The coded masks (second and third rows) always perform significantly better than the random masks used in the top row. Both the quadratic-residue MURA coded masks and the FRT-based coded masks have a comparably small SNR in their reconstructions, with error maps that both appear rather noise-like (less long range correlations) than for the corresponding random-mask reconstructions. This difference may be attributed to the fact that, while the average-subtracted intensity of the random masks is orthogonal in expectation value but not necessarily orthogonal \cite{CeddiaPaganin2018}, greater efficiency is achieved when the background-subtracted illuminating intensity maps are strictly orthogonal rather than merely orthogonal in expectation value \cite{GureyevGhost2018}.

\begin{figure}[!h]
\centering
\noindent
\includegraphics[width=0.4\linewidth]{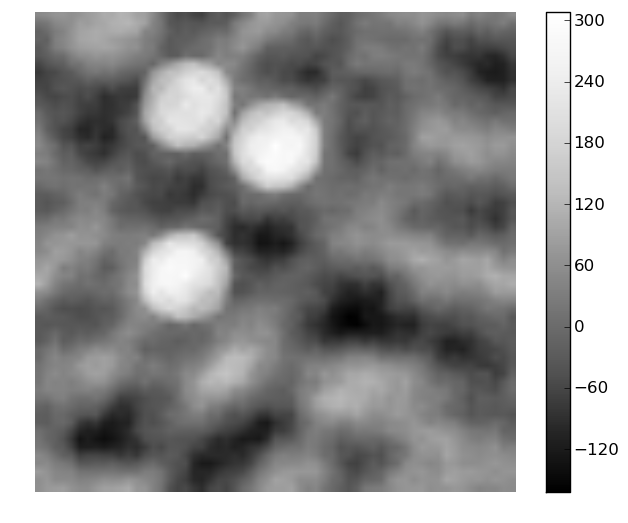}%
\includegraphics[width=0.4\linewidth]{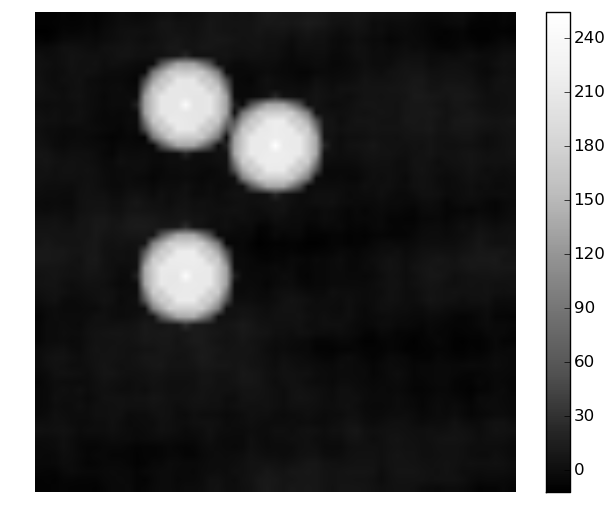}\\
\includegraphics[width=0.4\linewidth]{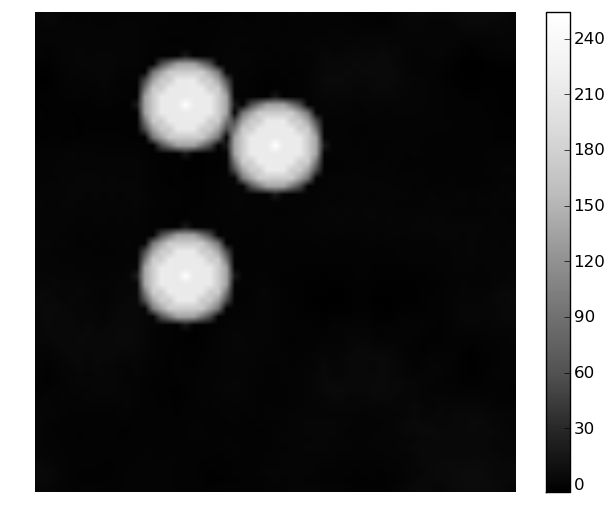}%
\includegraphics[width=0.4\linewidth]{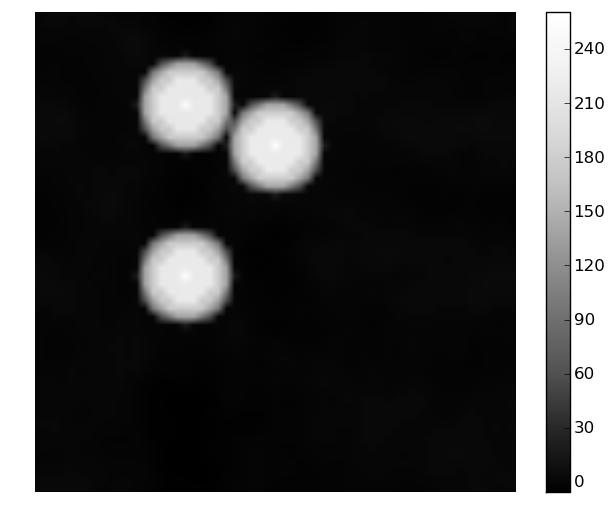}
\caption{
XC ghost image recovery for (T) random mask, (B:L--R) quadratic residue mask, finite Radon transform based mask. Recovery using XC from 3481 bucket measurements. (TR) Improvement in recovery for the random mask with 100 iterations of IXC. 
\label{fig:ghost_imaging_coded_apertures}\label{fig:iterative_ghost_imaging_coded_apertures}}
\end{figure}

Since these masks give an orthogonal basis when average-subtraction is applied to the resulting ensemble of intensity maps, no real benefit is gained from IXC.  In this instance, IXC only corrects for the slight variations shown in Fig. \ref{fig:coded_apertures}--R. The bucket value residuals and MAD from the known input do reduce, but the image quality is all but identical. Hence the associated images will not be displayed in this paper. 

Results from applying the IXC code to the {\em random} mask data is presented in Fig. \ref{fig:iterative_ghost_imaging_coded_apertures}. Note that (for reduced input data, i.e., $B < {\mathcal N}^2$) after IXC, both the residual and MAD is lower than that for the coded-apertures case. However, as can be seen in Figs. \ref{fig:ghost_imaging_coded_apertures} and \ref{fig:iterative_ghost_imaging_coded_apertures}, the errors include more low spatial-frequency components than that for the coded apertures. For the simulations presented here, then, it seems that there is a slight advantage in using specific coded-apertures over randomly generated apertures. This observation is consistent with the previously cited theoretical studies \cite{CeddiaPaganin2018, GureyevGhost2018}.

\subsubsection{Tomographic performance}

Our investigations regarding masks now pass from 2D X-ray GI to 3D X-ray GI.  The latter situation admits a choice to whether one utilizes the same set of masks for each object orientation, or a different set of masks for each object orientation. As we shall see, while the former situation is easier to realize experimentally, the latter gives superior reconstructions.

Simulated tomographic reconstructions were performed in two ways: (i) FBP from XC recovered projections (see Fig. \ref{fig:ghost_fbp_coded_apertures}), and (ii) 10 iterations of the XC-SIRT utilized earlier (see Fig. \ref{fig:ghost_sirt_coded_apertures}). Note that here  we have only included results for the FRT based mask (since the MURA mask gave similar results) and results from a random mask for comparison.

\begin{figure}[!h]
\centering
\noindent
\includegraphics[width=0.4\linewidth]{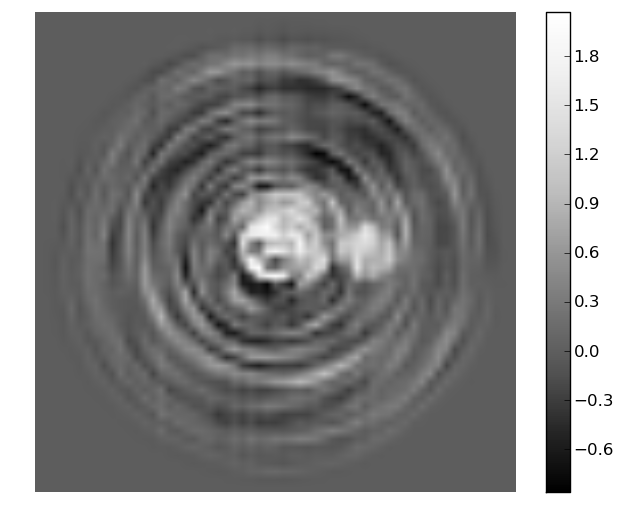}%
\includegraphics[width=0.4\linewidth]{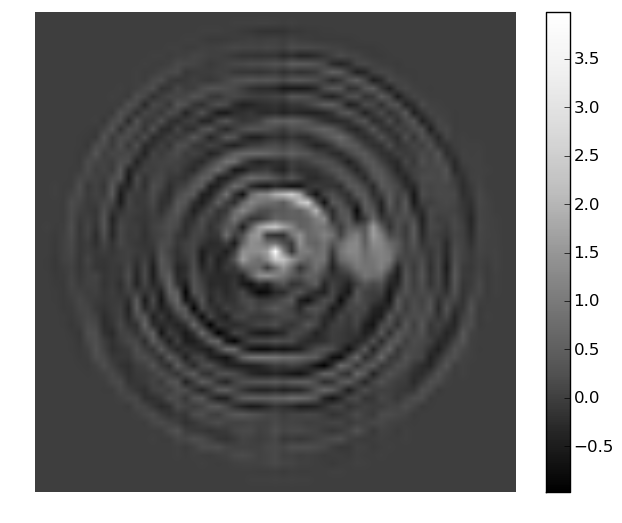}\\
\includegraphics[width=0.4\linewidth]{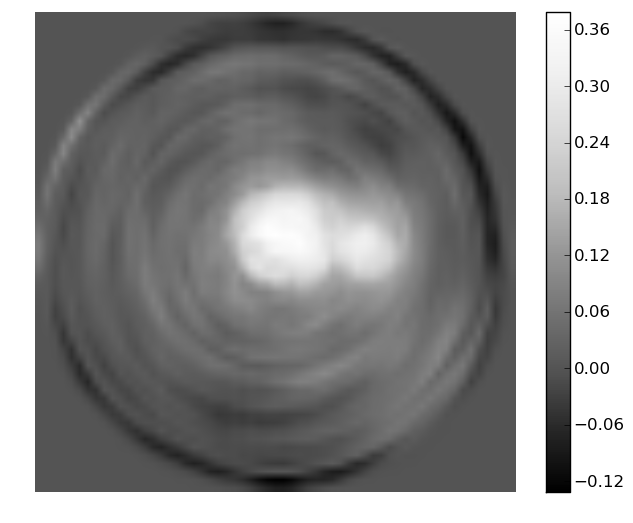}%
\includegraphics[width=0.4\linewidth]{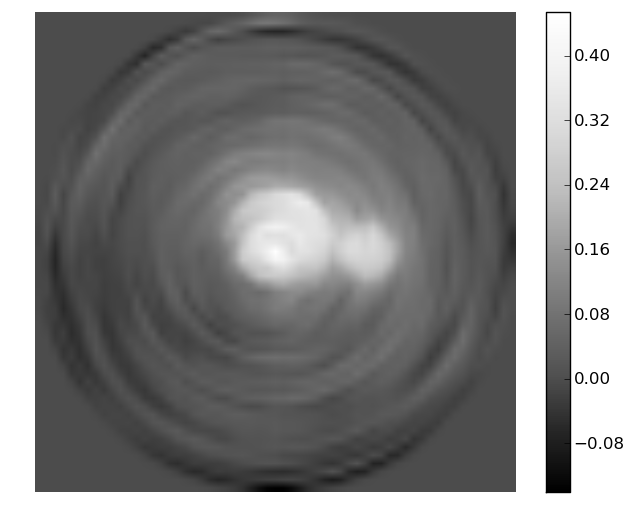}
\caption{
X-ray ghost tomography for (L-R) random mask and finite Radon transform based mask. Recovery using (T--B) FBP and 10 iterations of XC-SIRT from 1740 bucket measurements collected with the {\em same} random set of coded-aperture positions per viewing angle. 
\label{fig:ghost_fbp_coded_apertures}\label{fig:ghost_sirt_coded_apertures}}
\end{figure}

The main observation that can be made is the ring artifacts. The tomogram degradation due to these rings increases with reduced number of measurements. The artifacts have arisen since the simulated data has utilized the same set of mask positions for each projection angle. Apart from these artifacts we see that (as was the case for ghost imaging) the errors from the random mask have a lower spatial-frequency component polluting the reconstruction.  This may be compared to the reconstructions obtained using the perfect coded arrays, in which such low-freqency artifacts are absent. Moreover, we see in Fig. \ref{fig:ghost_sirt_coded_apertures} that XC-SIRT goes a little way towards cleaning up these errors and the ring artifacts in general.

\subsection{Ring artifacts}
\label{sect:ringartifacts}

To demonstrate the cause of the ring artifacts, we generated datasets with the same number of bucket measurements per viewing angle (as for the above demonstration), but with the mask positions selected {\em randomly}. Results are presented in Fig. \ref{fig:ghost_random_mask_coded_apertures}
for the FRT based mask using both the FBP and XC-SIRT reconstruction methods.  These simulations clearly suggest the use of a different set of masks for each object viewing angle as an effective means for  suppressing ring artifacts in X-ray ghost tomography.

\begin{figure}[!h]
\centering
\noindent
\includegraphics[width=0.4\linewidth]{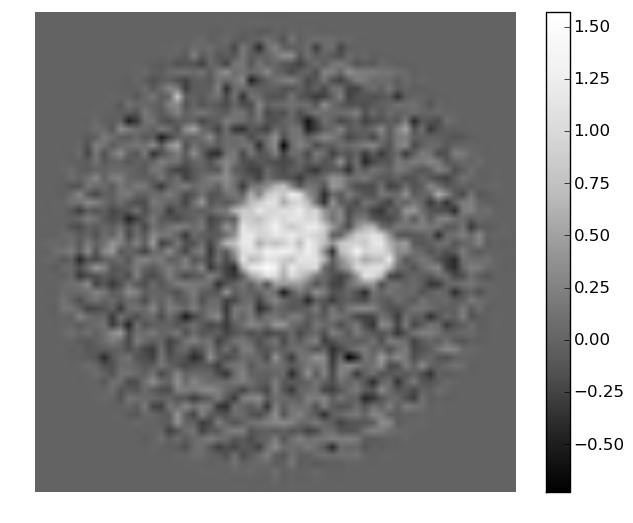}%
\includegraphics[width=0.4\linewidth]{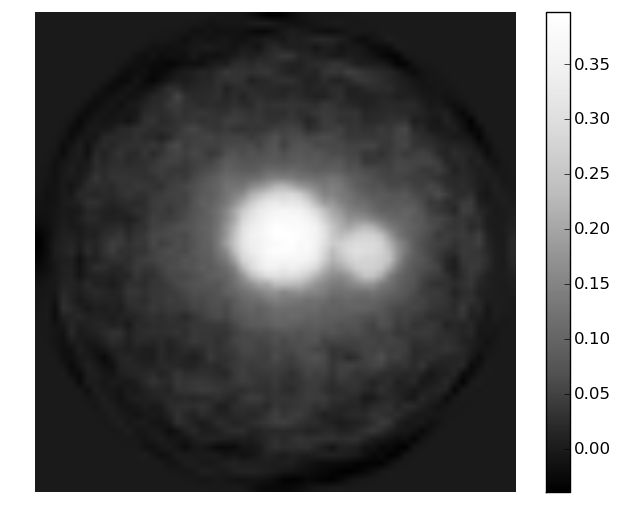}
\caption{
X-ray ghost tomography for finite Radon transform based mask. Recovery using (L--R) FBP, and 10 iterations of XC-SIRT from 1740 bucket measurements collected with a {\em different} random set of coded-aperture positions per viewing angle.  Ring artifacts are suppressed compared to the reconstructions obtained using the {\em same} set of masks for each viewing angle---{\em cf.} Figs~\ref{fig:ghost_fbp_coded_apertures} and \ref{fig:ghost_sirt_coded_apertures}}
\label{fig:ghost_random_mask_coded_apertures}
\end{figure}

To conclude, coded-apertures have slightly improved performance compared with random masks, owing to the background-subtracted intensities from the former being strictly orthogonal while the background-subtracted intensities from the latter are only orthogonal in expectation value \cite{CeddiaPaganin2018, GureyevGhost2018}. More important is the conclusion that use of a different set of illuminating masks for each object orientation significantly suppresses ring artifacts relative to utilizing the same set of masks for each object orientation.  Therefore, if one utilizes a single transversely-displaced mask for all measurements in tomographic X-ray GI, that mask should have a number of speckles (if random) or local intensity maxima (coded aperture) that is on the order of the total number of bucket measurements that will be taken. Stated more precisely, the single mask used for all bucket measurements should have at least $(MN)_{\textrm{max}}+\mathcal{N}^2$ speckles (if random) or local intensity maxima (coded aperture), where $(MN)_{\textrm{max}}$ is the maximum number of bucket measurements that will be taken, and $\mathcal{N}^2$ is the number of pixels required to be resolved in each 2D X-ray ghost projection.

\section{Discussion}
\label{sec:discuss}

X-ray ghost imaging is a field in its absolute infancy.  As of this writing, the authors are aware of only four published papers \cite{Yu2016,pelliccia2016experimental,Schori2017,zhang2017table} and one preprint \cite{pelliccia2017practical} reporting experimental realizations of X-ray ghost imaging.  To our knowledge, experimental X-ray ghost tomography has never been reported.  It is this gap in the literature that has inspired us to undertake the present study, which sketches how X-ray ghost imaging and ghost tomography may develop in the future.

One key question is whether 3D X-ray ghost imaging needs the intermediate step of a series of 2D X-ray ghost projections.  Our answer was ``not necessarily'', as we presented both indirect and direct methods to perform X-ray ghost tomography.  These were: (i) to obtain the 3D ghost tomogram via reconstructed 2D ghost projections, and (ii) to proceed directly from the X-ray ghost-tomography data to the 3D reconstruction.  It is not clear which approach will be more effective in future developments of the field. A virtue of the latter method its holistic approach in which one uses all of the data simultaneously, rather than first analyzing separate subsets of the data for each object orientation.  

We also  considered the choice of illuminating masks.  It is clearly advantageous---albeit experimentally more complex---to have a different illuminating mask pattern for each bucket signal obtained in an X-ray ghost tomography experiment,  We drew this conclusion based on the the ring artifacts that appear when one uses the same set of illuminating patterns for each object orientation.  In the longer-term development of the field of X-ray ghost tomography, the development of more sophisticated ring-artifact-removal algorithms may well alter this conclusion, particularly in contexts where {\em a priori} knowledge about the class of imaged samples may be employed.            

A further consideration, relating to both 2D and 3D X-ray ghost imaging, is the use of random versus coded masks.  In certain contexts, such as when using shot noise from individual charged-particle bunches to generate random speckle fields in a ghost imaging context \cite{pelliccia2016experimental}, one does not have a choice.  However, in many experimental circumstances one will have the choice between random and coded masks.  Which class of mask is superior?  We suspect that, in analogy with what is done with spatial light modulators, appropriately designed coded apertures will typically perform better than random masks, on account of the fact that the corresponding coded-aperture illumination patterns can be designed to be strictly orthogonal (after background subtraction, and subject of course to the usual uncertainties of experimental reality) \cite{GureyevGhost2018}, whereas the corresponding background-subtracted spatially random patterns only have the weaker property of being orthogonal in expectation value \cite{CeddiaPaganin2018}.         

One clear albeit unsurprising message of the current study is the fact that considerable efficiencies may be obtained in X-ray ghost tomography, by going beyond the ``vanilla'' XC ghost reconstruction formula given in Eq.~\ref{equation:XC_method}.  While this XC method for ghost imaging is conceptually appealing, since it can be trivially derived from first principles by considering the ensemble of background-subtracted illuminating fields to form an orthogonal (or approximately orthogonal) set \cite{pelliccia2017practical}, it is clearly much less efficient when compared with various iterative refinement methods such as iterative cross correlation, compressed sensing {\em etc.}  We suspect that, in the future, refinement methods based on machine learning and artificial intelligence \cite{kemp2017propagation, Rivenson2018} will become of progressively greater importance for X-ray ghost imaging, together of course with the strides being made in the burgeoning field of compressive sensing \cite{Qaisar2013}.  This will likely be a key avenue for future research, driven primarily by the quest for improved reconstructions using a minimal number of probe photons.

This leads us to the question of whether or not X-ray ghost imaging may enable reduced dose relative to competing protocols.  The answer at this stage is ``maybe'', with inequalities having been developed which, if violated, imply ghost imaging to enjoy reduced dose relative to its direct-imaging counterpart \cite{CeddiaPaganin2018, GureyevGhost2018}.  Indeed, the {\em logical possibility} that ghost imaging {\em may} reduce dose may be demonstrated by the following admittedly contrived example.  Suppose that one performs an X-ray ghost imaging experiment of a sample whose transmission function is a greyscale map of the Leonardo da Vinci's Mona Lisa, using a {\em single} illuminating intensity field that is also a greyscale map of the Mona Lisa.  The resulting reconstruction, which will be proportional to the illuminating intensity multiplied by the single bucket measurement, will give an excellent reconstruction of the object with minimal dose.  This contrived example is indicative of the more general result that X-ray GI may reduce dose, {\em e.g.} in cases where ``the class of imaged objects is strongly correlated with a small number of illumination patterns'' \cite{GureyevGhost2018}.  It remains unclear as to whether, in the longer term, X-ray GI can really reduce dose in realistic imaging contexts where sufficiently low doses are important ({\em e.g.} in the context of radiation damage to living tissues, radiation damage to other biological or radiation-sensitive materials, efficiently small acquisition time in an industrial-testing context, {\em etc}.).  Of relevance, in this context, is the ability of a low-resolution low-dose ghost reconstruction to first locate a region of interest (ROI) in a sample, so that subsequent illumination patterns can be adapted by confining them to this ROI \cite{Sun2016}.

\section{Future Research}
\label{sec:FutureWork}

The extent to which one can use polychromatic radiation, for X-ray ghost imaging, would also form an interesting avenue for future research.  This question is spurred by the fact that laboratory-based X-ray sources are typically polychromatic, and become severely limited in flux if too stringently monochromated.  Another motivation is given by the higher throughput permitted by not overly monochromating a synchrotron X-ray beam.  Can one work with polychromatic radiation in an X-ray ghost imaging setting?  An important fact, in this regard, is that nowhere in our development did we need to explicitly refer to the coherence of the beam. For example, if one is performing computational X-ray imaging with a laboratory source, as has already been reported experimentally \cite{Schori2017,zhang2017table}, one can work with a fully polychromatic beam.  For 3D X-ray ghost imaging in this context, one could first perform 2D ghost imaging for each orientation of the object, and then correct for the beam hardening \cite{brooks1976beam} (two step method). Beam-hardening effects can be modelled in the forward process of the one-step reconstruction, (that proceeds directly from measured data to the X-ray tomographic reconstruction), with an estimate of the polychromatic attenuation refined during iteration, {\it e.g.} \cite{krumm2008reducing, vangompel2011iterative}.

Another avenue for future work is prompted by the advances made in X-ray imaging in recent decades, due to the rise of X-ray phase contrast \cite{Wilkins2014}.  Through various incarnations including propagation-based X-ray phase contrast, analyzer-crystal phase contrast, grating-based phase contrast, speckle-tracking X-ray phase contrast {\em etc.} \cite{paganin2006coherent}, the harnessing of phase contrast in an X-ray imaging setting has led to significant increases in image contrast and resolution, as well as protocols for the (often rather substantial) reduction of sample dose \cite{SNRboost5}. The so-called Paganin method (PM) \cite{paganin2002} has been particularly successful in this regard, with acquisition-time-reductions in the tens of thousands being typical \cite{SNRboost5, SNRboost1, SNRboost2, SNRboost3, SNRboost4}.  While phase contrast has been incorporated into visible-light GI protocols through the use of interferometers \cite{shirai2011ghost, ghostphase1, ghostphase2, ghostphase3}, these are difficult to translate to an X-ray setting.  Indeed, the standard X-ray GI setup is totally insensitive to any refractive effects (phase shifts) imparted by the sample since such effects have no influence on the total number of photons registered in each bucket measurement.  A protocol has recently been proposed for X-ray phase contrast ghost imaging \cite{CeddiaPaganin2018}, which belongs to the class of SNR-boosting X-ray phase retrieval algorithms \cite{SNRboost5, gureyev2017unreasonable} epitomised by the previously-mentioned PM.  Such phase-contrast X-ray ghost imaging protocols, particularly in light of the SNR-boosting property, may form an interesting avenue for future developments of X-ray phase-contrast ghost tomography.  Such developments would also be relevant to the preceding discussion regarding dose reduction.

\section{Conclusion}

X-ray ghost tomography has been considered in this numerical study.  Particular focus was paid to denoising schemes, dose fractionation, and considerations regarding spatially-random versus coded-aperture masks.  We also studied both the origin and suppression of ring artefacts in X-ray ghost tomography.  Lastly, we gave some broad pointers for possible future avenues for research in the still-infant field of X-ray ghost imaging in general, and X-ray ghost tomography in particular.  

\label{sec:conclus}


\section*{Acknowledgment}

AMK and GRM acknowledge the financial support of the Australian Research Council and FEI-Thermo Fisher Scientific through Linkage Project LP150101040.  Financial support by the Experiment Division of the ESRF for DMP to visit in 2018 is gratefully acknowledged.   All authors thank Dr Woei M. (Steve) Lee and Dr Giuliano Scarcelli for assistance with the literature survey.  We acknowledge useful discussions with David Ceddia, Margie P. Olbinado, Timothy C. Petersen, Alexander Rack and Tapio P. Simula.

\ifCLASSOPTIONcaptionsoff
  \newpage
\fi



\bibliographystyle{IEEEtran}
\bibliography{IEEEabrv,references}
\end{document}